\documentclass[]{aa}
\usepackage[varg]{txfonts}

\usepackage{array, multirow, graphicx,natbib,txfonts,color,tabularx,amsmath}
\usepackage{lscape,float,placeins}
\usepackage[dvipsnames]{xcolor} \usepackage[colorlinks=true,allcolors=blue,urlcolor=CadetBlue,breaklinks]{hyperref}
\usepackage{afterpage}
\usepackage{threeparttable}
\bibpunct{(}{)}{;}{a}{}{,}

\usepackage{xspace}

\newcommand{\ang}{$\rm \AA$}

\newcommand{\XY}[2]{$\left[\mbox{#1/#2}\right]$}
\newcommand{\eqnXY}[2]{\left[\mbox{#1/#2}\right]}

\newcommand{\eqnFeH}{\eqnXY{Fe}{H}}

\newcommand{\eH}[1]{\XY{#1}{H}}
\newcommand{\eqneH}[1]{\eqnXY{#1}{H}}
\newcommand{\eqnalphaFe}{\eqnXY{$\alpha$}{Fe}}

\newcommand{\kms}{km\,s$^{-1}$}
\newcommand{\Teff}{$T_{\mbox{{\scriptsize eff}}}$}
\newcommand{\logg}{$\log g$}
\newcommand{\eqnTeff}{T_{\mbox{{\scriptsize eff}}}}
\newcommand{\eqnlogg}{\ensuremath{\log g}}

\newcommand{\vmic}{$v_{\mbox{{\scriptsize mic}}}$}
\newcommand{\eqnvmic}{v_{\mbox{{\scriptsize mic}}}}

\newcommand{\Abund}[1]{A(\text{#1})}
\newcommand{\Msol}{\text M_\odot}

\newcommand{\conf}[2]{$\rm #1 \, #2$}

\newcommand{\stagger}{\textsc{Stagger}\@\xspace}
\newcommand{\cobold}{{CO$^5$BOLD}\@\xspace}
\newcommand{\multitd}{\textsc{multi3d}\@\xspace}
\newcommand{\avgtd}{$\langle3\text{D}\rangle$\@\xspace}
\newcommand{\starfit}{\textsc{starfit}\@\xspace}
\newcommand{\thestar}{SMSS0313-6708\@\xspace}
\newcommand{\christliebstar}{HE\,0107-5240\@\xspace}

\newcommand*{\eg}{e.g.\@\xspace}
\newcommand*{\ie}{i.e.\@\xspace}

\usepackage{upgreek} \newcommand\micron{{\rm \upmu m}} 

\begin{document}

\title{3D NLTE analysis of the most iron-deficient star, SMSS0313-6708}

\author{T.~Nordlander\inst{\ref{upp1}} \and A.~M.~Amarsi\inst{\ref{anu}} \and K.~Lind\inst{\ref{upp1},\ref{mpia}} \and M.~Asplund\inst{\ref{anu}} \and P.~S.~Barklem\inst{\ref{upp2}} \and A.~R.~Casey\inst{\ref{cambridge}} \and R.~Collet\inst{\ref{anu},\ref{aarhus}} \and J.~Leenaarts\inst{\ref{stockholm}}}

\offprints{Thomas.Nordlander@physics.uu.se}

\institute{
        Division of Astronomy and Space Physics, Department of Physics and Astronomy, Uppsala University, Box 516, SE-751 20 Uppsala, Sweden \label{upp1} \label{upp2}
	\and Research School of Astronomy and Astrophysics, Australian National University, ACT 2611, Australia \label{anu}
	\and Max-Planck-Institut f\"ur Astronomie, K\"onigstuhl 17, 69117, Heidelberg, Germany \label{mpia}
	\and Institute of Astronomy, University of Cambridge, Madingley Road, Cambridge CB3 0HA, UK \label{cambridge}
	\and Stellar Astrophysics Centre, Department of Physics and Astronomy, Aarhus University, Ny Munkegade 120, DK-8000 Aarhus C, Denmark \label{aarhus}
	\and Institute for Solar Physics, Stockholm University, SE-106 91 Stockholm, Sweden \label{stockholm}
}

\date{Received  / Accepted}

\authorrunning{T. Nordlander \etalname}
\titlerunning{3D NLTE analysis of SMSS0313-6708}

\abstract
{Models of star formation in the early universe require a detailed understanding of accretion, fragmentation and radiative feedback in metal-free molecular clouds. Different simulations predict different initial mass functions of the first stars, ranging from predominantly low mass (0.1--10\,$\Msol$), to massive (10--100\,$\Msol$), or even supermassive (100--1000\,$\Msol$). The mass distribution of the first stars should lead to unique chemical imprints on the low-mass second and later generation metal-poor stars still in existence. The chemical composition of \thestar, which has the lowest abundances of Ca and Fe of any star known, indicates it was enriched by a single massive supernova. }
{The photospheres of metal-poor stars 
are relatively transparent in the UV, which may lead to large three-dimensional (3D) effects as well as departures from local thermodynamical equilibrium (LTE), even for weak spectral lines. 
If 3D effects and departures from LTE (NLTE) are ignored or treated incorrectly, errors in the inferred abundances may significantly bias the inferred properties of the polluting supernovae. 
We redetermine the chemical composition of \thestar by means of the most realistic methods available, and compare the results to predicted supernova yields.}
{We employ a 3D hydrodynamical \stagger model atmosphere and 3D NLTE radiative transfer to obtain accurate abundances for Li, Na, Mg, Al, Ca and Fe. The model atoms employ realistic collisional rates, with no calibrated free parameters. }
{We find significantly higher abundances in 3D NLTE than 1D LTE by 0.8\,dex for Fe, and 0.5\,dex for Mg, Al and Ca, while Li and Na are unaffected to within 0.03\,dex.
In particular, our upper limit for $[\text{Fe}/\text H]$ is now a factor ten larger, at $[\text{Fe}/\text H] < -6.53$ ($3 \sigma$), than previous estimates based on \avgtd NLTE (\ie using averaged 3D models). This higher estimate is due to a conservative upper limit estimation, updated NLTE data, and 3D$-$\avgtd NLTE differences, all of which lead to a higher abundance determination.}
{We find that supernova yields for models in a wide range of progenitor masses reproduce the revised chemical composition. In addition to massive progenitors of 20--60\,$\Msol$\ exploding with low energies (1--2\,B, where $1\,\text{B} = 10^{51}$\,erg), we also find good fits for progenitors of 10\,$\Msol$, with very low explosion energies ($< 1$\,B). We cannot reconcile the new abundances with supernovae or hypernovae with explosion energies above 2.5\,B, nor with pair-instability supernovae.
}

\keywords{radiative transfer -- stars: abundances -- Stars: Population III -- techniques: spectroscopic -- stars: individual: SMSS J031300.36 -- supernovae: general}

\maketitle


\section{Introduction}\label{sect:intro}
The chemical composition of extremely metal-poor stars offers unique insight into the early evolution of the Galaxy and the properties of the first stars, the so-called Population III.
No Population III star has ever been observed, and their IMF is debated due to the sensitivity of the models to the physics of accretion and fragmentation. 

Several groups have run hydrodynamical cosmological simulations coupled with primordial gas chemistry, where dark matter halos of $10^6\,\Msol$ collapse gravitationally to form the first stars at redshifts $z \approx 20$--30 \citep[\eg][]{gao_first_2007,bromm_formation_2013}.
\citet{abel_formation_2002} simulated such a minihalo, in which a self-gravitating dense core of $\sim$100\,$\Msol$ cooled by molecular hydrogen formed a single protostellar core. 
Depending on the accretion scenario, they argued that within a few thousand years after the simulations ended, the star would reach a mass of at least $30\,\Msol$, and possibly more than 100\,$\Msol$ within 1\,Myr. 
No other stars would form due to the immense radiation field of the young hot star, which would rapidly photodissociate all molecular hydrogen in the entire minihalo (pausing further star formation), and later disrupt it as the star explodes in a supernova after a few million years.

Refined simulations by \citet{greif_simulations_2011} highlight the importance of fragmentation in the accretion disk, leading to the formation of clusters of $\sim$10 protostars per minihalo. Their simulations indicated low protostellar masses (0.1--10\,$\Msol$) in a flat (top-heavy) distribution. The final mass of most of these objects would likely be higher, but the accretion phase of nearly half of the objects ended already at this stage when dynamical interactions ejected them from the star forming region, meaning long-lived low-mass Population III stars could very well still exist today.

\citet{hirano_one_2014} ran 2D axisymmetric radiation hydrodynamical stellar structure calculations, simulating the evolution of the central protostar in a self-collapsing gas cloud until the point where UV feedback shut off accretion and the star reached the main sequence. Their simulations resulted in a top-heavy stellar mass distribution with roughly equal numbers in the ranges 10--100 and 100--1000\,$\Msol$. Lower-mass stars were not produced, as their simulations did not consider accretion disk fragmentation. 
\citet{susa_mass_2014} reached similar conclusions from 3D simulations, with a mass spectrum covering 1--300\,$\Msol$, with stars on the lower end having been ejected from the star forming region.
The large number of $\sim$100\,$\Msol$ stars produced in both these simulations imply copious pair-instability supernovae should have preceded the formation of Population II stars. 
\citet{susa_mass_2014} concluded that the limited resolution of their simulations likely results in a bias toward higher stellar mass, potentially by 50--100\,\% on the upper end of the IMF. With the enormous numerical challenges of simulating the formation of the first stars, it is important to constrain theory by observations.

Despite several surveys specifically designed to discover extremely metal-poor objects in the halo \citep[\eg][]{beers1992,cayrel2001,christlieb2003,jacobson_high-resolution_2015} and bulge \citep{howes_extremely_2015}, no metal-free stars have been detected, and only a single star indicating pollution from pair-instability supernovae (\citealt{aoki_chemical_2014}, but see also \citealt{aoki_very_2015}) although this may be a selection effect \citep{karlsson_uncovering_2008}. 
A handful of extremely metal-poor stars have been identified to exhibit a chemical composition indicative of pollution from one or just a few core-collapse supernovae \citep[\eg][]{christlieb2002,frebel2005,caffau2011}, with \thestar\ being the most likely candidate to have been polluted by a single supernova \citep{keller2014,bessell2015}. The remarkable nearly line-free spectrum of \thestar indicating
extremely low abundances of calcium ($\eqneH{Ca} = -7.3$) and only upper limits to all heavier elements including iron ($\eqnFeH < -7.5$), but 
relatively high abundances of carbon and oxygen ($\eqneH C \approx \eqneH O \approx -2.5$) as well as magnesium ($\eqneH{Mg} \approx -4.1$), was found to match well the predicted yields from a massive supernova progenitor (40--60\,$\Msol$) and low explosion energy (1.5--1.8\,B, where $1\,\text B = 10^{51}\,\text{erg}$).

Such comparisons with supernova model yields depend critically on the detailed composition, where \eg the ratios of light to iron-peak elements and odd-even variations are the most sensitive to explosion energy and progenitor mass. \citet{bessell2015} found that in their analysis, the odd-even effect constrained the mass of the progenitor despite no odd elements beyond Li being detected, with the strongest constraints coming from the upper limits to N (prefering lower masses), and Na and Al (prefering higher masses). 
Supernova yields overall, and odd-even effects in particular, are however strongly model dependent \citep[see Fig.~5 of][]{nomoto_nucleosynthesis_2013}, \eg, due to differences in how explosions are initiated.
On the observational side, the detailed physics of spectral line formation causes deviations from LTE (this case is known as non-LTE, hereafter NLTE) which are known to significantly affect the Al resonance line strength. Furthermore, the interplay between effects on line formation from the steeper temperature structures of 3D model atmospheres and NLTE effects is largely unknown for all but smaller representative sets of stellar parameters \citep[\eg,][]{asplund_multi-level_2003,lind2013,klevas2016}, with the exceptions of \citet{sbordone2010} and \citet{amarsi2016} who computed 3D NLTE grids for lithium and oxygen, respectively.

In the first analysis of \thestar, \citet{keller2014} used both one-dimensional (1D) hydrostatic models \citep[ATLAS9 models from][]{castelli2004} and \avgtd temporally and horizontally averaged three-dimensional (3D) hydrodynamical models \citep[from the \stagger grid,][]{magic2013_grid,magic2013_avg}, with NLTE line formation for Li, Na, Mg, Ca and Fe. 
The analysis was improved by \citet{bessell2015}, who analyzed molecular lines using 3D LTE radiative transfer, which is appropriate and necessary due to their very strong temperature dependence \citep{asplund_oh_2001}.

The fundamental shortcoming of 1D models is that they are hydrostatic and thus must implement convective energy transfer using a parametrized mechanism, typically mixing-length theory \citep{bohm-vitense_uber_1958}. 
Additionally, 1D models use a free local velocity parameter called microturbulence (\vmic), representing small-scale velocity fields which serve as a line desaturation mechanism \citep{asplund_line_2000}.

While \avgtd temporally and horizontally averaged hydrodynamical models, are more realistic than their 1D hydrostatic counterparts, 
the averaging process loses physical properties of the 3D atmospheric structure including surface granulation (\ie horizontal inhomogeneities) and local velocity fields. 
The importance of surface granulation has been demonstrated in the past \citep[\eg][]{collet_three-dimensional_2007,magic2013_avg,dobrovolskas2013}, where significant effects were found in LTE abundance analyses of metal-poor red giant stars.
\citet{magic2013_avg} found 3D$-$1D abundance corrections for weak low-excitation \ion{Fe}i lines of $-0.3$\,dex, and 3D$-$\avgtd LTE abundance corrections of $-0.1$\,dex, indicating that the effects of surface granulation and the steeper temperature gradient have similar magnitude.
In contrast, \citet{dobrovolskas2013} found abundance corrections for weak resonance lines of \ion{Fe}i to be $-0.5$\,dex for both 3D$-$1D LTE and 3D$-$\avgtd LTE, indicating that the effects of granulation entirely dominate over those of the steeper temperature gradient.
Using the same models and averaging techniques as \citet{dobrovolskas2013}, \citet{klevas2016} compared 3D NLTE calculations to \avgtd NLTE and 1D NLTE calculations for \ion{Li}i, finding similar line strengths in 3D NLTE and 1D NLTE, in line with previous results \citep{asplund_multi-level_2003}. They find that as 3D LTE line formation is inappropriate for \ion{Li}i, the NLTE and 3D effects cannot simply be added together.

We present here an abundance analysis of \thestar employing NLTE line formation in a 3D hydrodynamical model atmosphere, for Li, Na, Mg, Al, Ca and Fe, using state-of-the-art atomic models with accurate collisional data. 
The models of Al and Fe in particular use recent accurate transition rates for hydrogen collisions, which have previously been a major source of uncertainty due to the dependence on classical models with empirical scaling factors. We introduce the spectroscopic analysis, the hydrodynamical models and the 3D NLTE modeling in Sect.~\ref{sect:methods}, present the 3D NLTE effects and the resulting abundances in Sect.~\ref{sect:results}, and compare the resulting abundance determinations (Sect.~\ref{sect:abundances}) to supernova yield models and discuss the inferred properties of the progenitor and explosion in Sect.~\ref{sect:discussion}.



\section{Methods}\label{sect:methods}

We have determined the abundances of Li, Na, Mg, Al, Ca and Fe with state-of-the-art 3D NLTE modeling using the \multitd\ code \citep{leenaarts_multi3d_2009}, with updates as described by \citet{amarsi2016}. 
We present
the observational data in Section~\ref{sect:obs},
introduce the 3D model atmosphere in Section~\ref{sect:atmosphere},
and give details on the 3D NLTE problem in Section~\ref{sect:nlte} and our model atoms in Section~\ref{sect:atoms}.

\subsection{Observations and spectroscopic analysis}\label{sect:obs}
We use existing spectra of \thestar\ taken with the MIKE spectrograph \citep{bernstein2003} on the 6.5\,m Magellan Clay Telescope and UVES \citep{dekker2000} on the VLT. 
The UVES spectrum covers the UV region 3060--3860\,\AA\ and the optical region 4790--6800\,\AA\ at resolving power $R \approx 41\,000$ \citep[see][for details]{bessell2015}. 
We extracted pipeline-reduced data from the ESO archive, and coadded the 31 exposures after applying radial velocity shifts found by cross-correlation matching. The resulting spectrum has $S/N \approx 100$ in the region near 3800\,\AA, and $S/N \approx 200$--300 in the visual region.
The resonance lines of \ion{Al}i and \ion{Ca}{ii} are not available in these spectra, and we therefore analyze these lines using the MIKE spectrum \citep[see][]{keller2014} which has $R \approx 28\,000$ and $S/N \approx 80$ in this region.

For our abundance determination of lithium, magnesium and calcium, we fit the observed line profiles. 
Our 3D NLTE analysis is based on interpolated synthetic 3D NLTE spectra computed in a range of abundances, and we convolve the synthetic spectra with a fitted Gaussian representing the instrumental profile and rotational broadening. 
For the 1D modeling, this broadening also represents large-scale (macroturbulent) velocity fields. Uncertainties are based on the $\chi^2_\text{red}$ statistic, where we estimate the $S/N$ from the local continuum and typically use three free parameters (abundance, broadening and radial velocity). In the 1D abundance analysis, we use a fixed value of $\eqnvmic = 2\,$\kms.
For lithium, we use the \ion{Li}i 6707\,\AA\ resonance line, taking into account fine structure.
For magnesium, we use the two \ion{Mg}i UV lines at 3829 and 3832\,\AA\ as well as the \ion{Mg}ib 5167--5183\,\AA\ triplet. We rejected the strongest UV line at 3838\,\AA, where the continuum placement is significantly distorted by the nearby Balmer H9 line at 3835\,\AA. 
For calcium, we use the unblended \ion{Ca}{ii} 3933\,\AA\ line. For comparison, we have also determined abundances from the 3968\,\AA\ line, which is suppressed by the nearby Balmer H$\varepsilon$ line at 3970\,\AA.

For sodium, aluminium and iron, we do not detect any lines and report only upper limits. 
We use equivalent width uncertainties estimated using Cayrel's formula \citep{Cayrel1988,Cayrel2004}, $\sigma_\text{EW} = \frac{1.5}{S/N} \sqrt{\text{FWHM} * \delta x}$, where $\delta x$ is the pixel sampling. We also propagate representative uncertainties in the continuum placement into the final equivalent width uncertainty.
For sodium, we use the \ion{Na}iD 5895\,\AA\ line which is less affected by interstellar absorption than the neighbouring \ion{Na}iD 5889\,\AA. 
For aluminium, we use the unblended \ion{Al}i line at 3961\,\AA. 
For iron, we stack spectra in the vicinity of the unblended \ion{Fe}i lines at 3440.6, 3581.2, 3719.9, 3737.1, 3820.4 and 3859.9\,\AA. We estimate $S/N$ from the nearby continuum of the stacked spectrum.

\subsection{Model atmosphere} \label{sect:atmosphere}
We adopt the hydrodynamical model computed by \citet{bessell2015} using the \stagger\ code \citep{nordlund1995,stein_simulations_1998}, with $\eqnTeff = 5150\,\text K$, $\eqnlogg = 2.2$, $\eqnFeH = -5$.
The model was computed in a cartesian geometry with $240^3$ grid points representing a $1145^2\,\times\,458$\,Mm$^3$ box in the photosphere, using tailored opacities with standard alpha enhancement ($\eqnalphaFe = 0.4$) and enhanced CNO abundances ($\eqneH C = \eqneH N = -3.0$ and $\eqneH O = -3.6$). This composition was originally tailored to \christliebstar\ \citep{christlieb2004}, but is sufficiently close to the actual chemical composition of \thestar. 
The simulation used 12 opacity bins selected as a function of wavelength and formation height, and treats continuum scattering using the approximation described by \citet{collet_three-dimensional_2011}.
We adopt the abundances derived by \citet{bessell2015} to compute the equation of state and background continuous opacities and generate synthetic spectra. 
We use the reference solar abundances from \citet{asplund2009}.

For comparison purposes, we also perform the abundance analysis using a temporally and horizontally averaged one-dimensional representation of the hydrodynamical model, denoted \avgtd, as well as a 1D hydrostatic MARCS model \citep{gustafsson2008} computed with the same abundances by K. Eriksson. 
The \avgtd\ model used in this work has been averaged on surfaces of equal continuum optical depth $\tau_{500}$, but other methods exist in the literature \citep[see][]{magic2013_avg}.
We use $\eqnvmic = 2$\,\kms, and discuss this choice in Sects.~\ref{sect:flux} and \ref{sect:abundances}.

\begin{figure}
 \centerline{\includegraphics[width=9cm]{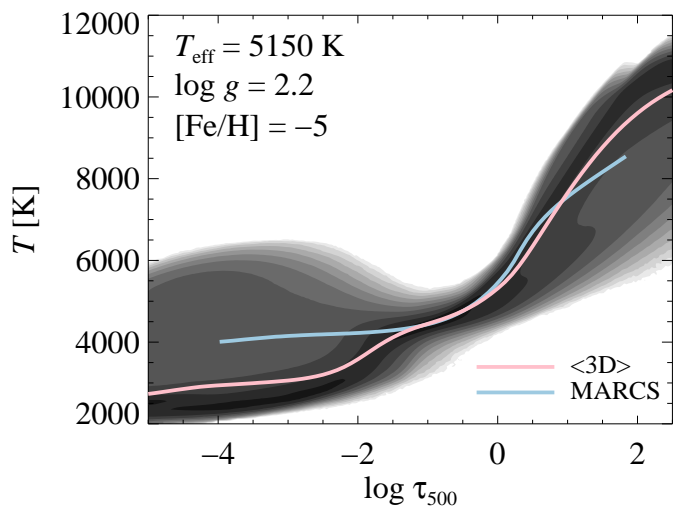}}
 \caption{Temperature stratification in the hydrodynamical model shown in a gray-scale histogram as a function of the continuum optical depth. 
Also shown are its spatial and temporal average (pink line), compared to a hydrostatic (1D) MARCS model with similar stellar parameters (blue line).
} \label{fig:Tstructure}
\end{figure}

The temperature distribution in the hydrodynamical model is compared to the MARCS model in Fig.~\ref{fig:Tstructure}. The outer layers ($\log \tau_{500} < -2$) of the hydrodynamical model are predominantly cooler than the MARCS model, due to the importance of adiabatic cooling by gas expansion \citep{asplund_3d_1999}. However, a significant fraction of the surface exhibits temperature inversion due to radiative reheating and mechanical compression above the intergranular lanes \citep[see \eg][]{rutten2004,magic2013_avg,delacruz2015}.

\begin{table*}
\caption{Abundance sensitivity to changes in stellar parameters} \label{tbl:sensitivity}
\centering
\begin{tabular}{l rr rr rr rr}
\hline\hline \noalign{\smallskip}
Element & \multicolumn2c{$\eqnTeff + 100$\,K} & \multicolumn2c{$\eqnlogg + 0.2$\,dex} & \multicolumn2c{$\eqnvmic + 0.5$\,\kms} & \multicolumn2c{Standard CNO\tablefootmark a} \\
    & NLTE & LTE & NLTE & LTE & NLTE & LTE & NLTE & LTE \\
\hline \noalign{\smallskip}
Li 		& $+0.10$ & $+0.10$ &  	$-0.01$ & $0.00$ &  	$0.00$ & $0.00$ &  	$-0.01$ & $-0.04$\\
Na  	& $+0.08$ & $+0.09$ &  	$ 0.00$ & $0.00$ &  	$0.00$ & $0.00$ &  	$-0.01$ & $-0.03$ \\
Mg  	& $+0.10$ & $+0.08$ &  	$-0.03$ & $-0.01$ &  	$-0.04$ & $-0.06$ &  	$-0.01$ & $-0.03$ \\
Al   	& $+0.11$ & $+0.10$ &  	$-0.02$ & $-0.01$ &  	$-0.02$ & $-0.02$ &  	$ 0.00$ & $-0.03$ \\
Ca 	& $+0.09$ & $+0.06$ &  	$+0.05$ & $+0.08$ &  	$-0.18$ & $-0.20$ &  	$-0.03$ & $-0.05$ \\
Fe   	& $+0.15$ & $+0.14$ &         $-0.05$ & $-0.01$ &     $0.00$ & $0.00$ &          $+0.07$ & $-0.05$ \\
\hline
\end{tabular}
\tablefoot{
	The sensitivities have been computed using interpolated standard-composition MARCS models, assuming $\eqnvmic = 2$\,\kms.
	\tablefoottext{a}{Abundances inferred from an atmospheric model computed with standard abundances, compared to our tailored model which uses $\eqneH C = \eqneH N = -3.0$ and $\eqneH O = -3.6$. The difference is given as $\text{standard} - \text{tailored}$.}
}
\end{table*}

We estimate the effect of changes in stellar parameters on the abundance determination using MARCS models in Table~\ref{tbl:sensitivity}. There, we also show the influence of using model atmospheres computed with non-standard CNO abundances, compared to the standard solar-scaled composition. As this has a small effect on the abundance determination, we judge that the difference in composition between that adopted for the model atmosphere calculation and the actual composition of the star likely does not significantly affect results.

To lessen the computational cost, we select representative snapshots from the hydrodynamical simulation sequence, and reduce these in size.
We found that we could reduce the model atmospheres in size from the original $240^3$ grid points to retain only every fourth column and interpolating to 100 horizontal layers, giving a total of $60^2 \times 100$ grid points. In the following, we shall always retain 100 horizontal layers and refer to models by their number of columns, $N_x^2$. 
The detailed numerical tests discussed in Sect.~\ref{sect:numerics} show that these simplifications introduce negligible systematic effects, of at most 0.03\,dex. 
The small temporal variations in strength of all lines investigated here ($\sigma \lesssim 0.02$\,dex) indicate that a set of just five snapshots is sufficient.

\subsection{3D NLTE calculations} \label{sect:nlte}

NLTE radiative transfer for late-type stars involves solving the equation of statistical equilibrium to compute NLTE atomic populations, while typically neglecting feedback on the model atmosphere. Details are given in several textbooks \citep[\eg][]{rutten2003,hubeny_theory_2014} and illustrative examples for cool stars are found in several reviews \citep[\eg][]{asplund_new_2005,bergemann_nlte_2014}. 
A limiting case is found in the deep photosphere, where the mean free path of photons is short compared to the gradient of local physical parameters, thus resulting in atomic populations close to LTE. In the high photosphere or chromosphere, densities are sufficiently low that non-local radiation drives higher transition rates than collisions, thus resulting in NLTE atomic populations. Spectral lines form in the intermediate layers, where radiative and collisional rates compete, and thus results are sensitive to the precise prescriptions for both. 

LTE radiative transfer in 3D model atmospheres is nowadays accessible even for large grids of calculations \citep[\eg][]{allende2013,magic2013_avg,magic2014_fe}, but only in a few cases feasible in NLTE \citep[\eg][]{asplund_multi-level_2003,sbordone2010,lind2013,steffen_photospheric_2015,klevas2016,amarsi2016}. For comparison, while the grid of LTE spectra computed by \citet{allende2013} took roughly one CPU-year to compute, we exceed that when solving the statistical equilibrium of iron at a single abundance in a single 3D model snapshot. 
The computation time for 3D NLTE calculations scales like $N_x N_y N_z N_\text{rays} N_\nu N_\text{iterations}$, with $N_{x,y,z}$ the geometrical dimensions of the atmospheric model and $N_\nu$ the number of frequencies. 
Here, we solve the statistical equilibrium using the 3D radiation field with $N_\text{rays} = 24$, distributed according to Carlson's quadrature set A4 \citep{carlson_numerical_1963,bruls_computing_1999}. 
Additionally, several snapshots from the hydrodynamical simulation are required to sample the temporal evolution.
All in all, 3D NLTE calculations turn out to be some $10^5$ times more expensive than in 1D for a given model atom.
In order to make these calculations feasible, care must thus be taken to reduce the complexity of both model atoms and atmospheres. 

Implementation details on the version of \multitd\ used here are given by \citet{leenaarts_multi3d_2009} and \citet{amarsi2016}. We note that the code implements isotropic coherent scattering in background processes, considering Rayleigh scattering by \ion H i in the far red wing of Lyman-$\alpha$ and by $\text{H}_2$, as well as Thomson scattering on electrons. Background line opacities, mainly due to hydrogen, were included as true absorption.

\subsection{Model atoms} \label{sect:atoms}

\begin{table}
\caption{Model atoms for NLTE calculations}
\label{tbl:atoms}
\centering
\begin{tabular}{l l l r r c}
\hline\hline
Species & $N_\text{levels}$\tablefootmark{a} & $N_\text{lines}$\tablefootmark{a} & $N_\text{b-f}$\tablefootmark{b} & $N_\nu$\tablefootmark{c} & Source \\
\hline
\ion{Li}{i, ii}	& 20, 1	& 91		& 20 & 1400  & 1, 2 \\
\ion{Na}{i, ii}	& 20, 1	& 91		& 20 & 1400  & 2, 3 \\
\ion{Mg}{i, ii, iii}	& 43, 2, 1	& 70, 1    & 8  & 1600  & * \\
\ion{Al}{i, ii}	& 42, 2	& 90		& 4  & 2100  & 4, * \\
\ion{Ca}{ii, iii}	& 5, 1		& 5		& 0  & 550   & 2, * \\
\ion{Fe}{i, ii, iii} 
	& 421, 12, 1	& 776, 12 	& 33 & 12000 & 5, * \\
\hline 
\hline
\end{tabular}
\tablefoot{
The ground states of \ion{Mg}{iii}, \ion{Ca}{iii} and \ion{Fe}{iii} are not significantly populated, but are included for technical reasons. \\
\tablefoottext{a}{The number of levels or lines in each listed ionization stage. }
\tablefoottext{b}{Number of radiative bound-free continuous transitions from the neutral to the singly ionized state. We do not consider bound-free transitions from the ionized states.}
\tablefoottext{c}{Approximate number of frequency points used to compute the radiation field when solving the statistical equilibrium.}
}
\tablebib{
(*) This work;
(1) \citet{lind2009};
(2) \citet{lind2013};
(3) \citet{lind2011};
(4) Nordlander et al. (in prep.);
(5) Lind et al. (in prep.), \citet{amarsi_non-lte_2016-1}
}
\end{table}

\begin{figure*}
\newcommand{\figscale}{0.24}

\newcommand\lcut{0} 
\newcommand\ltrim{3} 
\newcommand\rcut{1} 
\newcommand\bcut{2.3}
\newcommand\tcut{3em}
\centerline{
	\includegraphics[height=\figscale\textheight,clip,trim=\lcut em \bcut em \rcut em \tcut]{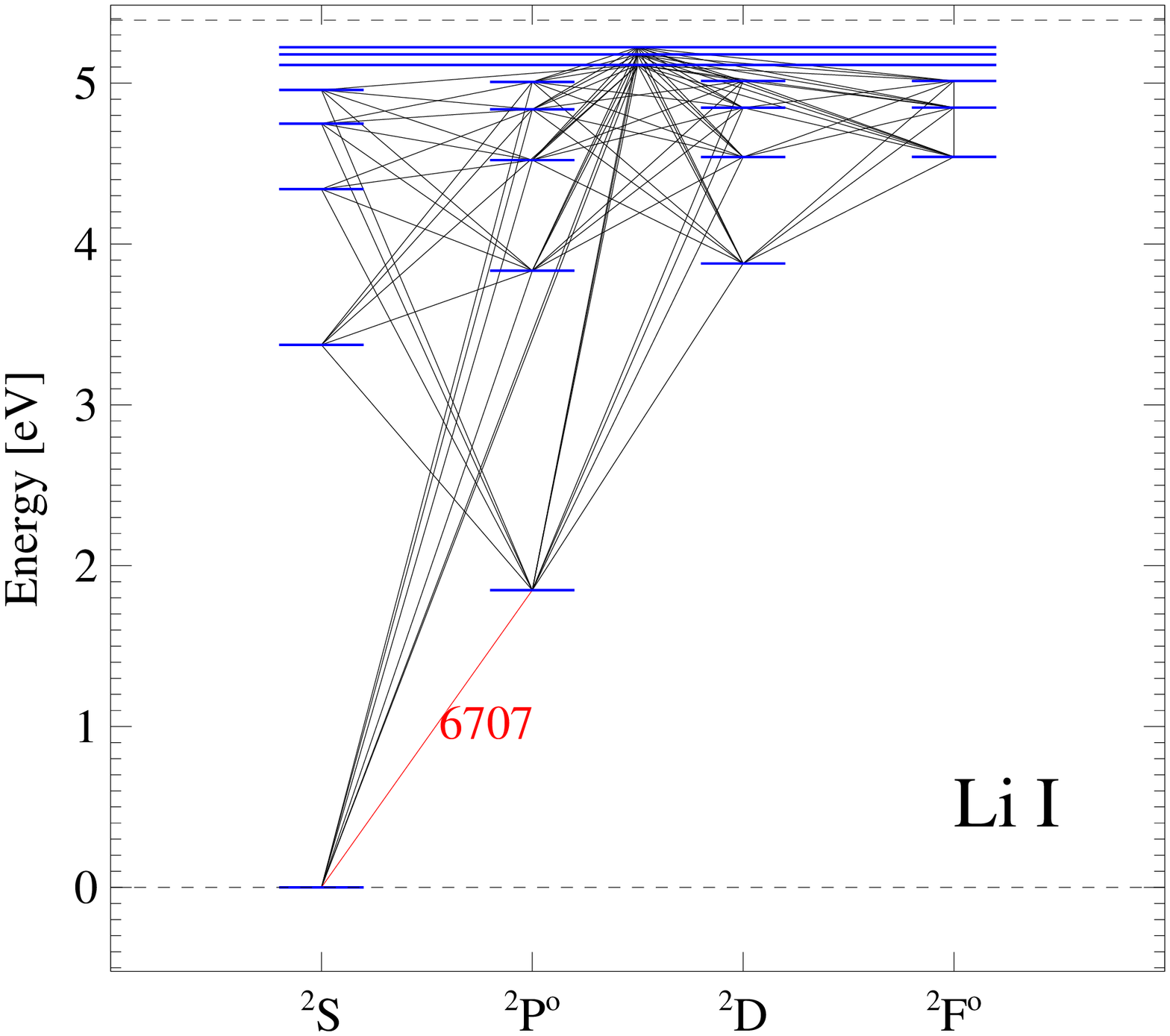}
	\includegraphics[height=\figscale\textheight,clip,trim=\ltrim em \bcut em \rcut em \tcut]{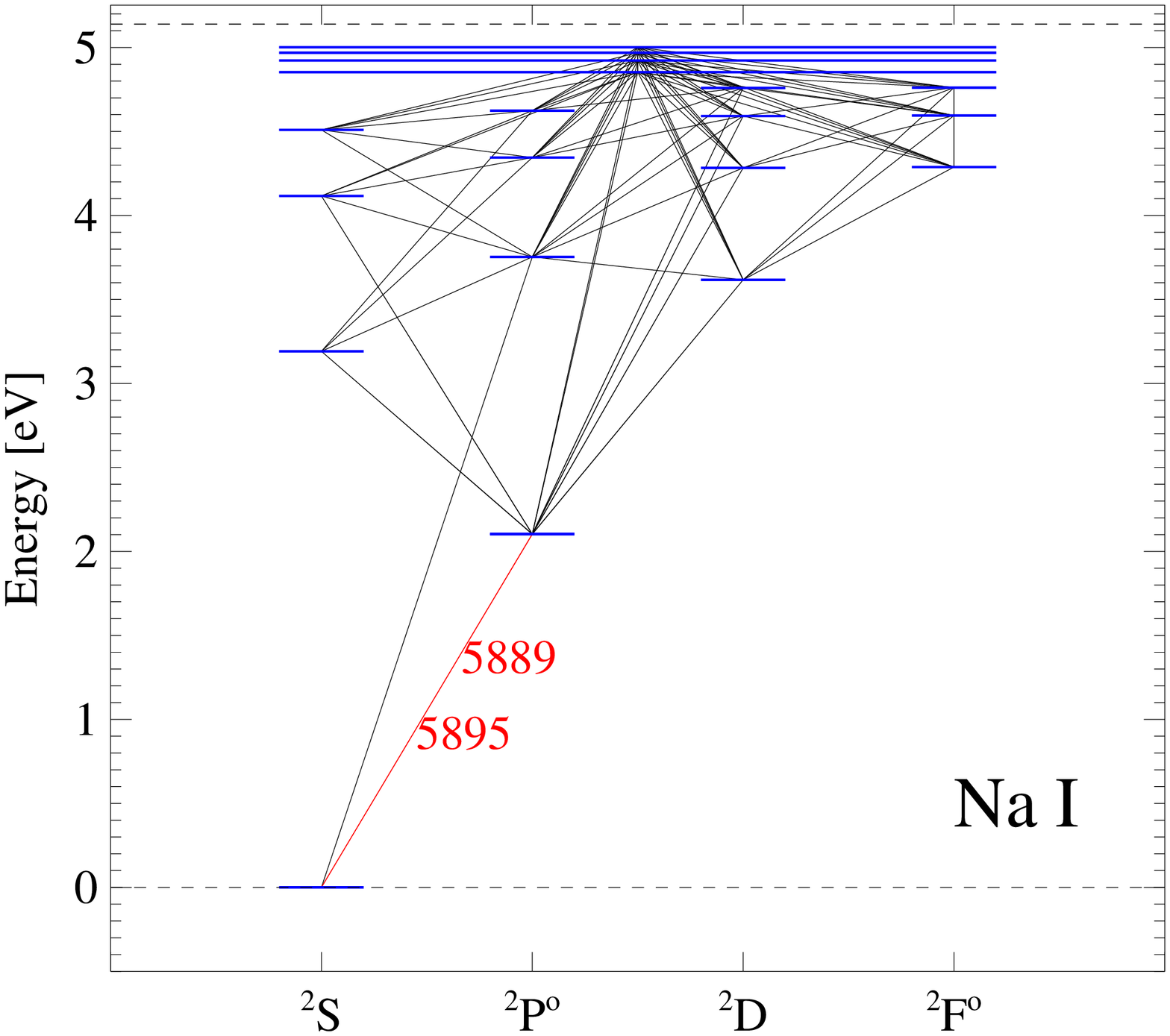} 
	\includegraphics[height=\figscale\textheight,clip,trim=\ltrim em \bcut em \rcut em \tcut]{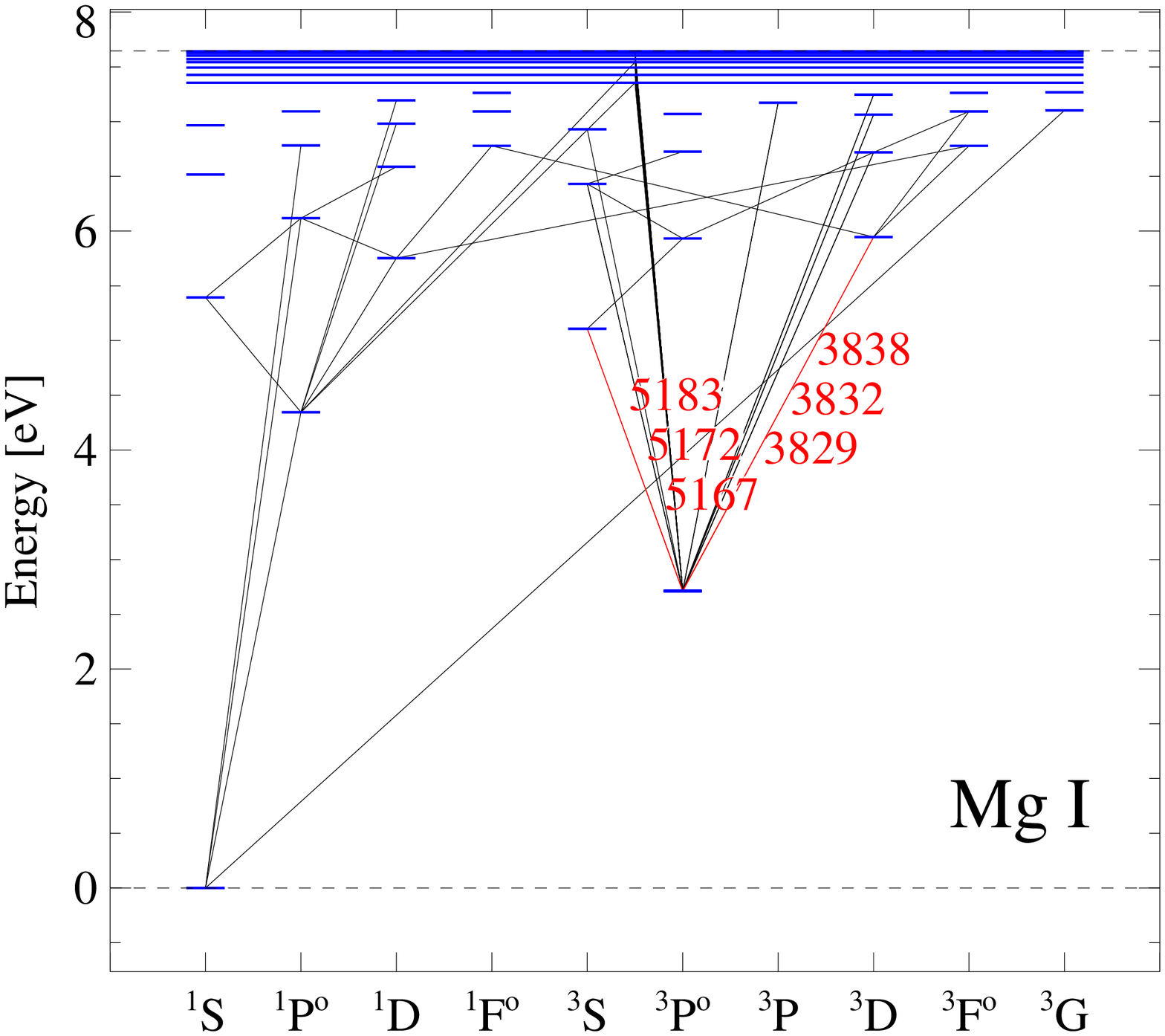} 
}
\centerline{
	\includegraphics[height=\figscale\textheight,clip,trim=\lcut em \bcut em \rcut em \tcut]{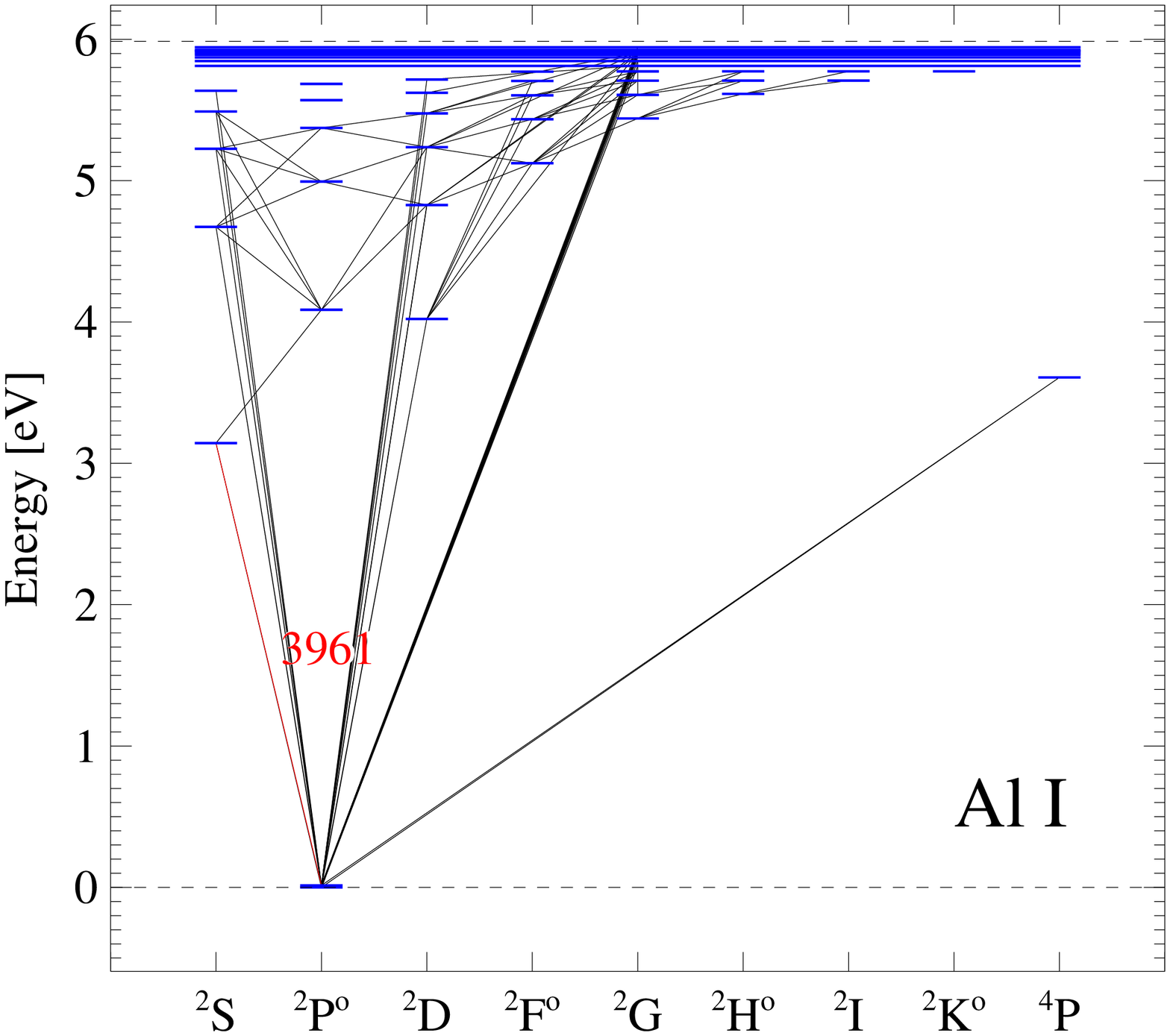} 
	\includegraphics[height=\figscale\textheight,clip,trim=\ltrim em \bcut em \rcut em \tcut]{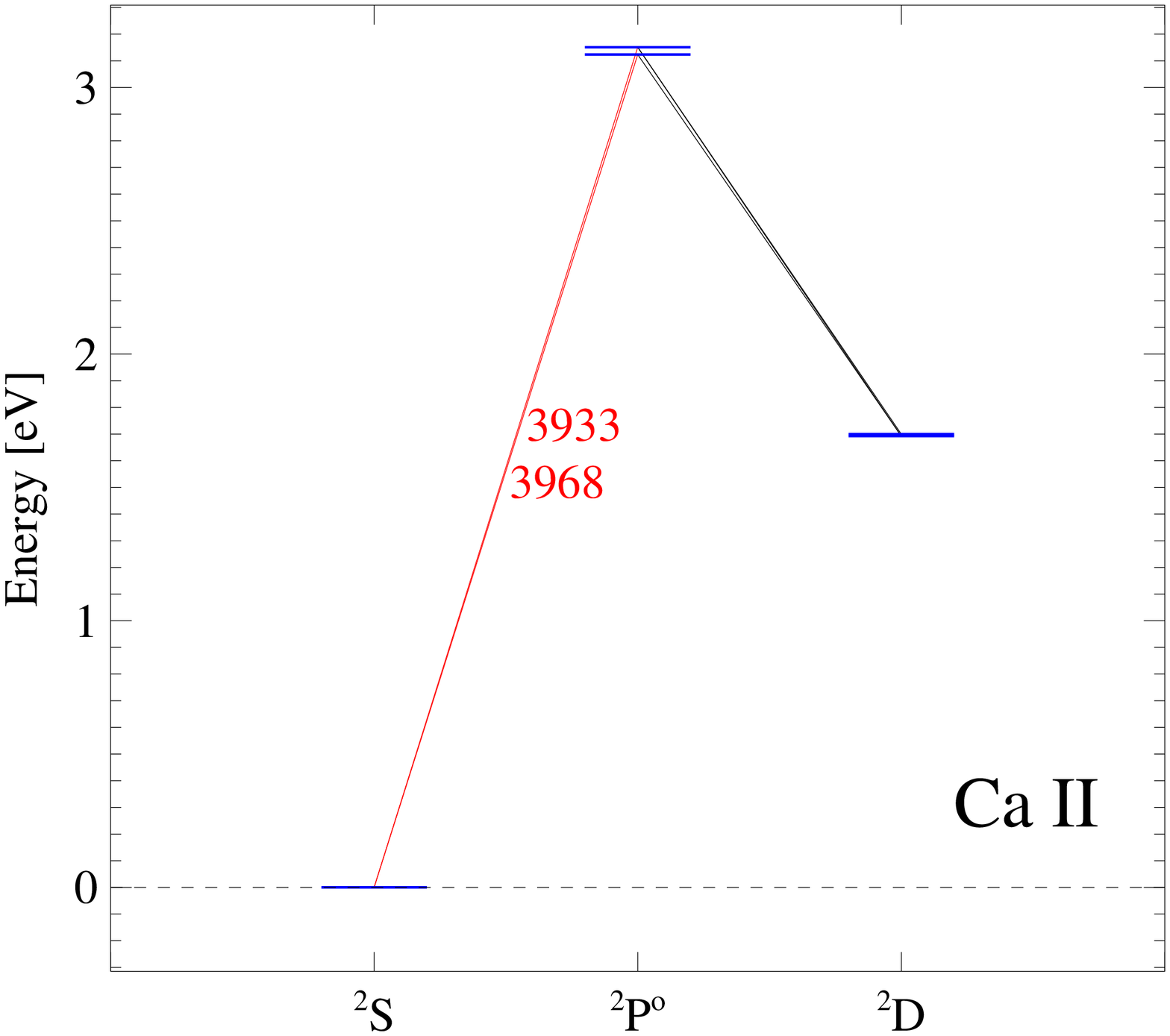} 
	\includegraphics[height=\figscale\textheight,clip,trim=\ltrim em \bcut em \rcut em \tcut]{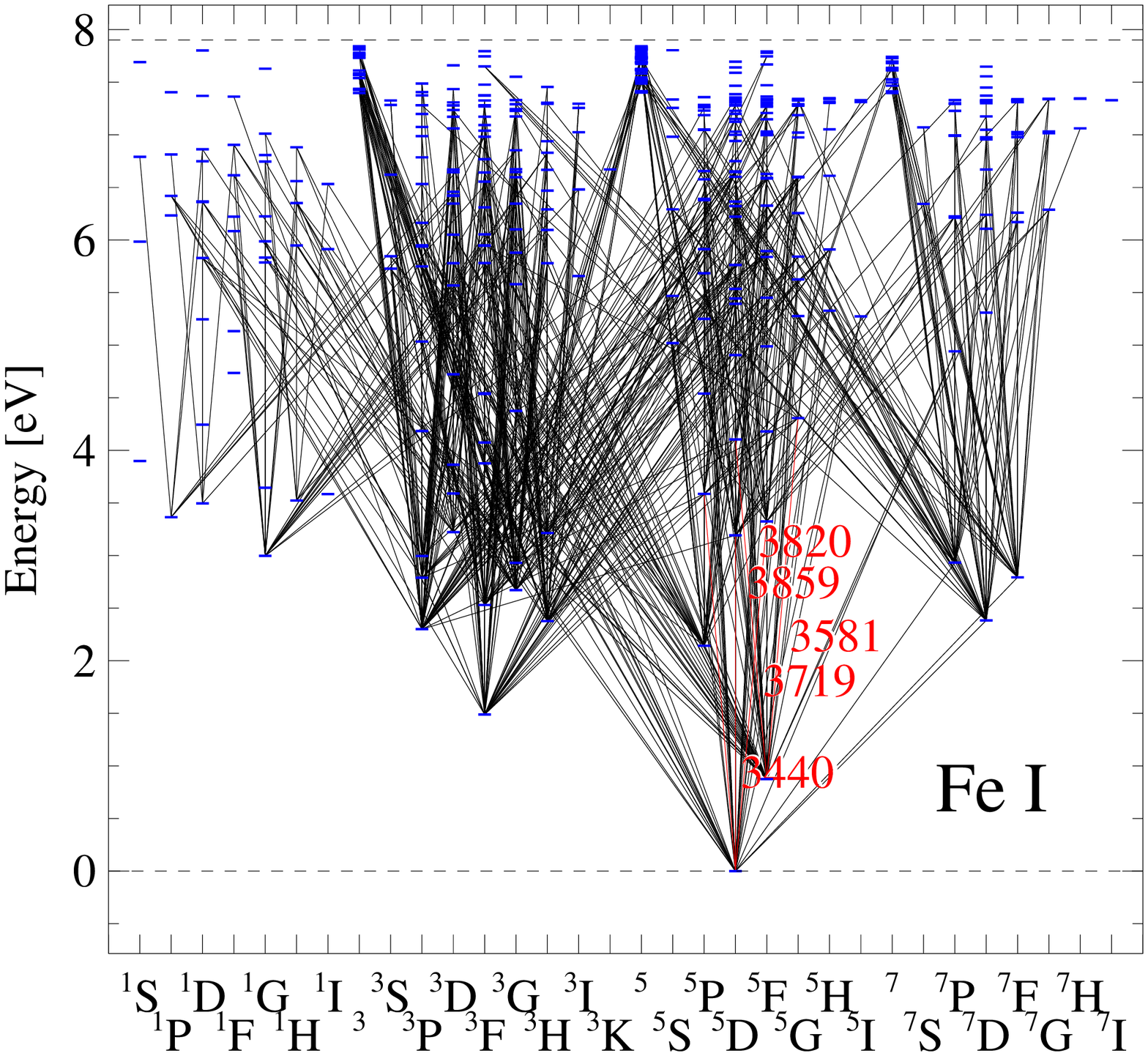} 
}
\caption{Term diagrams of the atoms analyzed in this work, illustrating all bound-bound transitions considered in the solution of the statistical equilibrium. Thicker lines are due to overlapping transitions. Lines or multiplets used in the abundance analysis are shown in red and labeled. The excited states of \ion{Mg}{ii}, \ion{Al}{ii} and \ion{Fe}{ii} are omitted from the figure.}
\label{fig:termdiags}
\end{figure*}

Due to the high computational cost of 3D NLTE calculations, the very detailed general-purpose atoms with (tens of) thousands of spectral lines typically used in 1D analyses must be modified before use. 
In order to lessen the computational cost, we reduce the number of radiative transitions by discarding those which do not have signficant impact on the statistical equilibrium, using an approach described by Lind et al. (in prep.). 
In summary, we quantify the relative importance of radiative transitions by comparing their radiative bracket, $|n_i r_{ij} - n_j r_{ji}|$, where $r_{ij}$ are radiative rate coefficients, and removing those transitions where this value (essentially the influence on the statistical equilibrium) is relatively small. We perform this test at $\tau_{500} = 0.1$ using a representative \avgtd\ \stagger\ model, requiring that NLTE populations of states involved in the lines we investigate here remain unaffected to within 2\,\% (0.01\,dex) at depths $\log \tau > -2$, or farther out in the case of stronger lines forming in the UV. We also require that the inferred abundance from all lines used in the analysis remain unaffected to within 0.01\,dex. 
We apply the same criterion when determining which atomic levels to retain, and always compare the final atom to the original one.
The complexity of our resulting model atoms is summarized in Table~\ref{tbl:atoms} and the corresponding term diagrams are shown in Fig.~\ref{fig:termdiags}. 
In particular, we note that we were able to discard all bound-free transitions of the ionized species, as these did not significantly affect the populations of the relevant atomic levels, even in \ion{Ca}{ii}. We could also significantly reduce the number of bound-free transitions in the neutral species of atoms where just a few of these were dominant.

Collisional transition rates for impact by electrons and hydrogen atoms are a major source of uncertainty for NLTE modeling.
Neutral hydrogen atoms are orders of magnitude more numerous than free electrons, but the low mass of electrons leads to high thermal velocities and thus large impact rates on atoms. Hydrogen atoms have low impact rates as they are relatively massive and thus slow, and according to the Massey criterion \citep{massey1949} this also leads to the expectation of adiabatic collisions with smaller inelastic collisional cross-sections. 
As calculations of accurate hydrogen collisional rates have become feasible, it is apparent that both hydrogen and electron collisions are important, and that classical recipes describing hydrogen collisions \citep{drawin1968} are typically far worse than order-of-magnitude estimates \citep[see][for recent reviews on the subject]{barklem_inelastic_2011,barklem_accurate_2016}.
These inaccuracies have traditionally been to some degree compensated for by introducing empirical scaling factors, which vary significantly from element to element as well as between authors \citep[see \eg Table~2 of][]{bergemann_nlte_2014}.
The model atoms used in this work use quantum mechanical calculations for hydrogen transition rates, and are free from empirical scaling factors.

\subsubsection{Lithium and sodium}
We use the same atoms as \citet{lind2013}, meaning slightly modified versions of those presented by \citet{lind2009,lind2011}. The Na atom is truncated at $n=10$, leaving 20 levels of \ion{Na}i, and for both atoms we consider only the 91 most important radiative transitions.
Hydrogen collision transition rates \citep{barklem_li,barklem_na} are based on quantum scattering calculations based on quantum chemical data for the LiH and NaH molecules \citep{belyaev_li,belyaev_na}. For lithium, the dominant electron collisional rates come from \citet{park1971}. While the R-matrix calculations performed by \citet{osorio_li} differed significantly, the dominant transitions happen to agree sufficiently well that the effects on the resonance line strength are smaller than 0.01\,dex.

\subsubsection{Magnesium}
We use a simplified but realistic Mg model atom assembled for this work. We adopt hydrogen collision rates from \citet{barklem_mg}, based on quantum mechanical cross-sections \citep{guitou_inelastic_2011,belyaev_mg} for low-lying states of \ion{Mg}{I}, based on quantum chemical data for the MgH molecule from \citet{guitou_accurate_2010}. For Rydberg transitions we computed rates based on the free electron model of \citet{kaulakys1985,kaulakys1991} using the implementation of \citet{paul_barklem_2016_50217}. The relative importance of electron and hydrogen collisional transitions was demonstrated by \citet{osorio2015}, indicating the strong sensitivity to hydrogen collisional rates in giants. We adopt energy levels, f-values, and photoionization cross-sections from TOPbase \citep{cunto_topbase_1993,butler_atomic_1993,mendoza_atomic_1995}, except for the \ion{Mg}ib lines where we adopt the laboratory f-values from \citet{aldenius2007}. The TOPbase photoionization cross-sections are tabulated using a dense frequency sampling necessary to resolve narrow resonance features. However, limitations in the calculations result in minor uncertainties in the positions (but not necessarily strengths) of resonances, which may spuriously coincide with narrow background opacity features. To reduce the potential bias of such coincidences, \citet{bautista_rap} suggested smoothing photoionization cross-sections to represent the inherent uncertainty of resonance energies. \citet{allende_rap} executed such smoothing, and we use their data, interpolated onto an equidistant frequency grid in order to sample the frequency dependence of background opacities.

Our Mg atom is represented by 43 levels of \ion{Mg}{I}, 2 levels of \ion{Mg}{II} and the \ion{Mg}{III} continuum. For \ion{Mg}{I}, we retain fine structure in the \conf{3p}{^3P^o} term, and $l$-splitting when $l \le 3$ and $n \le 6$. Levels with a given principal quantum number $n$ and $l \ge 4$ or $n > 6$ were combined, regardless of spin. In line with \citet{osorio2015}, we found that the collisional couplings were sensitive to the inclusion of high-lying states, and thus include states all the way up to $n=80$ in a coarse set of superlevels representing states with $n=9$--10, 11--12, 13--14, 15--20, 21--30, 31--50, 51--80. For \ion{Mg}i, we include 70 spectral lines, 47 of which originate from the \conf{3p}{^3P^o} states, and continua coupling the \ion{Mg}{II} ground state to the \conf{3p}{^3P^o} and \conf{3p}{^1P^o} states with photoionization edges at 2512 and 3756\,\AA\ (driving overionization), and five higher excited states with edges at 2.2\,$\micron$\ and beyond (driving over-recombination). The two levels of \ion{Mg}{II}, \conf{3s}{^2S} and \conf{3p}{^2P^o}, are collisionally and radiatively coupled via the resonance line at 2799\,\AA. In total, the bound-bound and bound-free radiative transitions are computed at 1600 frequency points.

\subsubsection{Aluminium}
We use a model atom to be described by Nordlander et al. (in prep.), with hydrogen collisional rates from \citet{belyaev_al_2} computed for low-lying states using a model approach based on semi-empirical couplings and the Landau-Zener model \citep{belyaev_al_1}, and rates computed following \citet{kaulakys1985,kaulakys1991} for Rydberg states. Electron collisional rates come from open-ADAS\footnote{ADF04 data products for \ion{Al}i and \ion{Al}{ii} computed in 2012, available online at \url{http//open.adas.ac.uk}} \citep[computed as described by][]{badnell2011} when possible, including forbidden transitions, and are otherwise computed using \citet{seaton1962} for allowed transitions.
Energy levels and f-values are taken from NIST \citep{kelleher2008} when possible, or otherwise the most recent data from Kurucz \citep{kurucz1995}. Photoionization cross-sections are again taken from \citet{allende_rap}, based on data from TOPbase \citep{mendoza_atomic_1995}. 

Our Al atom is represented by 42 levels of \ion{Al}{I} and 2 levels of \ion{Al}{II}. For \ion{Al}{I}, we resolve fine structure in the \conf{3p}{^2P^o} ground state, and we retain $l$-splitting when $n \le 8$, and otherwise represent the states by a single superlevel at a given $n$-value when $9 \le n \le 15$. States with $16 \le n \le 20$ are represented by a single superlevel. 
Levels of \ion{Al}{I} are coupled radiatively by 90 b-b transitions, 40 of which are resonance lines. 
We include also four bound-free transitions, coupling the fine-structure \conf{3p}{^2P^o} ground states and the \conf{3d}{^2D} and \conf{4p}{^2P^o} states to the \ion{Al}{II} continuum with ionization edges at 2075, 6312 and 6528\,\AA, driving overionization. 
The bound-bound and bound-free transitions use a total of 2100 frequency points. 
\ion{Al}{II} is represented by the ground and first excited state, with only collisional coupling.

\subsubsection{Calcium}
We modified the model atom of \citet{lind2013}. Only lines of \ion{Ca}{ii} are visible in the spectrum of \thestar, and as the dominant ionization stage, the behavior of \ion{Ca}{II} does not noticeably depend on \ion{Ca}{I}. 
We thus retain only the \conf{4s}{^2S} ground state of \ion{Ca}{II} plus the first two excited states (\conf{3d}{^2D} and \conf{4p}{^2P^o}, with resolved fine-structure) and the \ion{Ca}{III} continuum.
Electron collisional rates are based on R-matrix calculations \citep{melendez2007}, but the statistical equilibrium is not very sensitive to collisional rates.
We use energy levels from TOPbase (HE Saraph \& PJ Storey, to be published) and f-values from Kurucz \citep{kurucz1995} which are very similar to the data in NIST. 

We verified that NLTE effects on \ion{Ca}i indeed do not significantly affect the lines of \ion{Ca}{ii} on the level of 0.01\,dex, by running tests on a complete atom representing \ion{Ca}i by 39 levels, 244 b-b transitions and 39 b-f transitions.
For these tests, we included hydrogen collisions for \ion{Ca}i based on \citet{drawin1968} rescaled by $S_\text H = 0.1$ \citep[following][]{mashonkina_non-lte_2007}, as \citet{lind2013} found acceptable fulfillment of the excitation and ionization equilibria using these collisional rates for a set of very metal-poor dwarfs and subgiants. We note that \citet{barklem_excitation_2016-1} recently computed rates for charge transfer and collisional excitation of \ion{Ca}i, but do not expect these transition rates to affect the conclusion that the statistical equilibrium of \ion{Ca}{ii} is not significantly affected by NLTE effects on \ion{Ca}i.

We model the resonance \ion{Ca}{ii} H and K lines at 3933--3968\,\AA\ and the \ion{Ca}{ii} IR triplet at 8500\,\AA, using 550 frequency points. We neglect continuous transitions coupling \ion{Ca}{ii} to the \ion{Ca}{iii} continuum as the photoionization cross sections are rather small, with edges at $\lambda < 1420$\,\AA, resulting in negligible influence on the statistical equilibrium.

\subsubsection{Iron}

We utilize a heavily reduced version of the newly compiled model atom which will be described in an upcoming paper (Lind et al., in prep.), and has previously been applied by \citet{amarsi_non-lte_2016-1}.
This model atom adopts collision rates for Fe+H collisions calculated with the asymptotic two-electron method presented by \citet{barklem_excitation_2016-1} and applied there to Ca+H. The calculation used here includes 138 states of \ion{Fe}i, and 11 cores of \ion{Fe}{ii}, leading to the consideration of 17 symmetries of the FeH molecule.
This removes the systematic uncertainty of classical collisional rates \citep{drawin1968}, which in previous works were rescaled by a factor $S_\text H$ calibrated on standard stars.
Electron collisional rates were computed using the formula from \citet{vanregemorter1962}.
The statistical equilibrium of iron represents a formidable problem, as the atom is vastly more complex than the others investigated in this work. Even disregarding fine structure splitting, the full model atom considers $10^5$ radiative transitions, requiring sampling at $10^6$ frequency points, at an estimated computational cost of several millennia of CPU time for the full 3D NLTE solution. 

Our reduced atom is represented by 421 levels of \ion{Fe}{I}, 12 levels of \ion{Fe}{II}, and the \ion{Fe}{III} continuum. We have reduced the number of \ion{Fe}{II} levels to retain only the nine low-lying states required to correctly compute the partition function, as well as the upper states of a few resonance multiplets in the UV (\conf{z}{^6P^o}, \conf{z}{^6D^o} and \conf{z}{^6F^o}). Additionally, we do not resolve fine structure when solving the statistical equilibrium, and include only 776 and 12 spectral lines for \ion{Fe}{I} and \ion{Fe}{II}, respectively, as well as 33 continuous transitions coupling \ion{Fe}{I} to \ion{Fe}{II}. In total, these radiative transitions are computed at 12000 frequencies.



\section{Results}\label{sect:results}
We discuss the NLTE mechanisms and the resulting 3D NLTE effects in Sect.~\ref{sect:mu0} and in Sect.~\ref{sect:flux} where we also compare to 1D modeling. We estimate the errors in Sect.~\ref{sect:numerics}, and present the results of the abundance analysis in Sect.~\ref{sect:abundances}.

\subsection{Resolved disk-center 3D NLTE effects} \label{sect:mu0}

\begin{figure*}
\centerline{
	\includegraphics[clip,trim=0 3em 0 0]{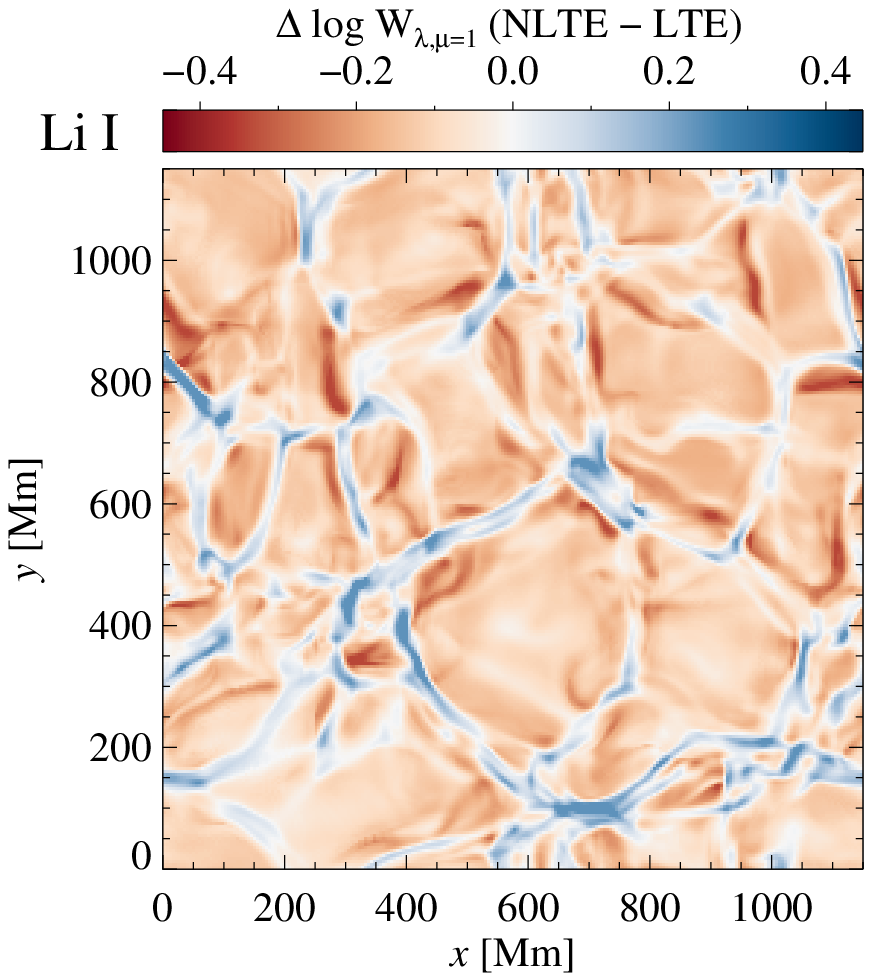} 
	\includegraphics[clip,trim=1.2em 3em 0 0]{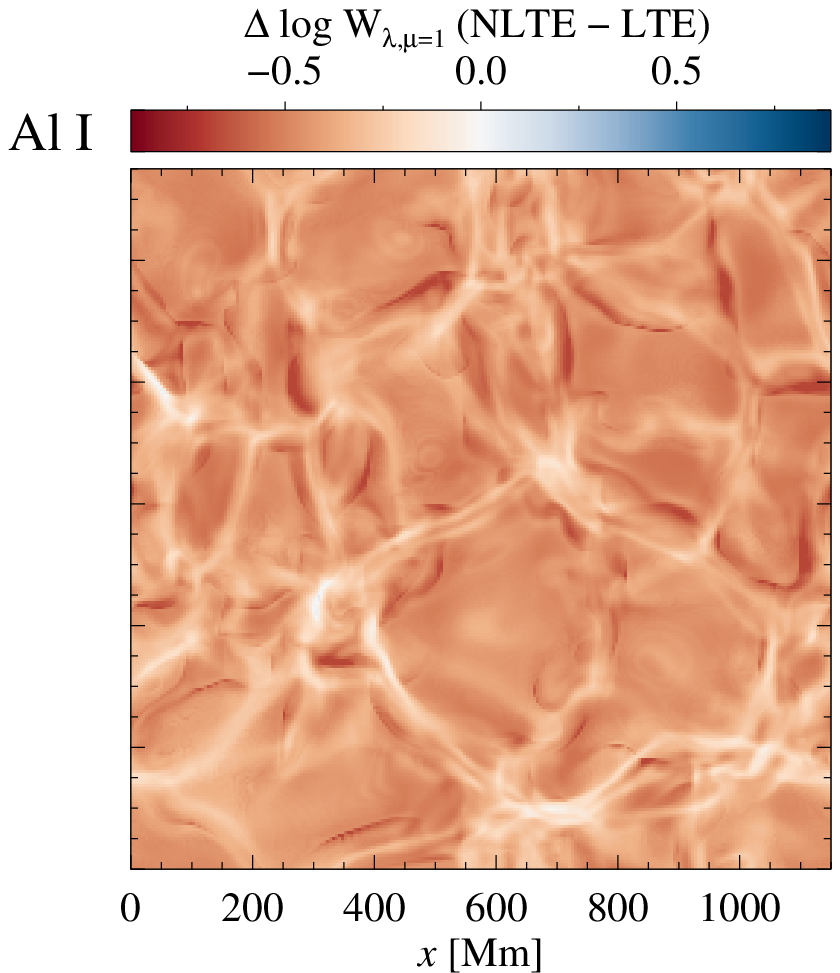}
} \vspace{.5em}
\centerline{
	\includegraphics[clip,trim=0 0 0 0]{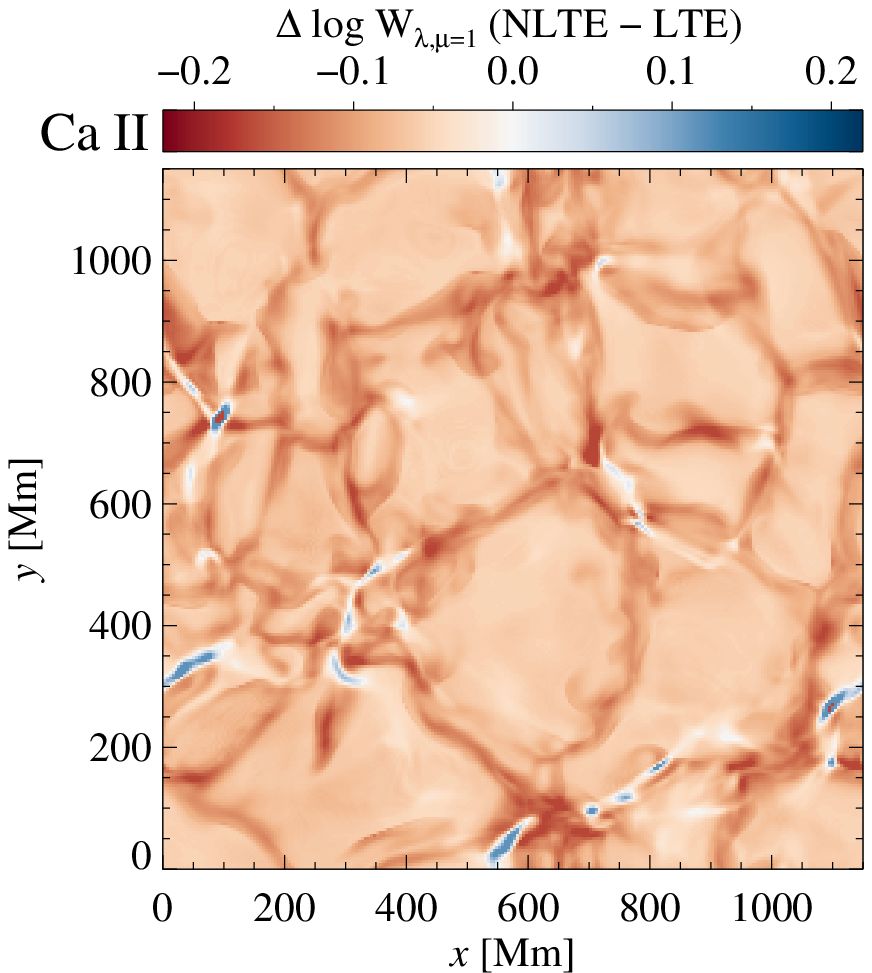} 
	\includegraphics[clip,trim=1.2em 0 0 0]{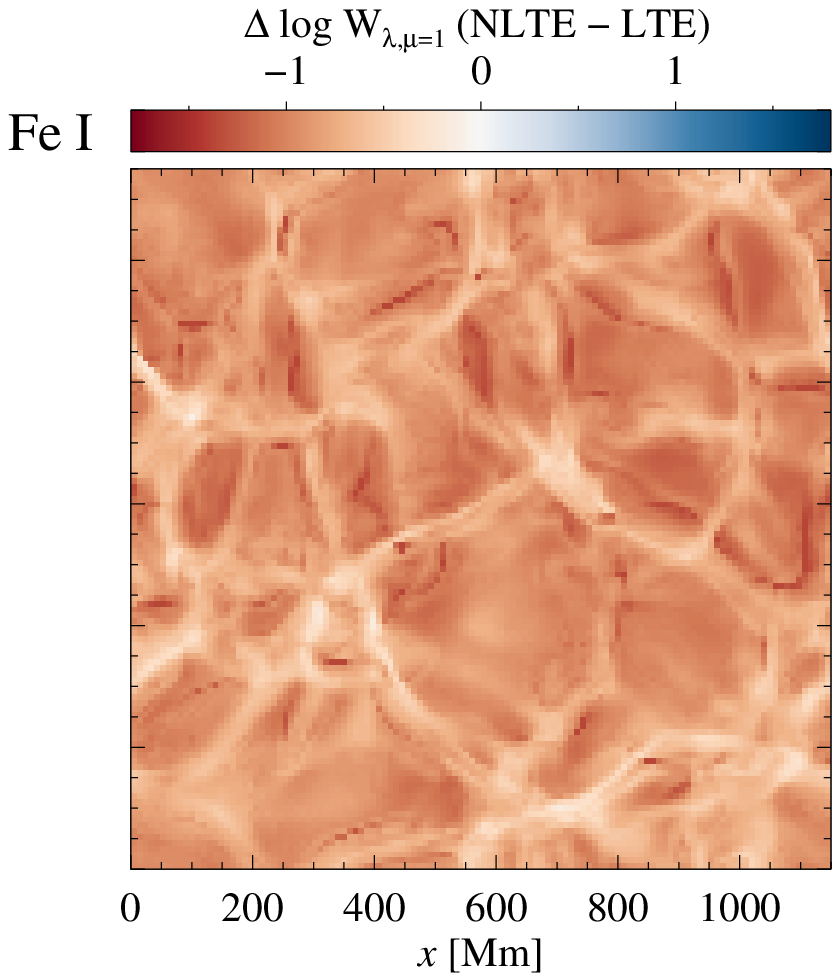} 
}
\caption{Spatially resolved NLTE equivalent width correction at disk-center intensity ($\mu = 1$), shown for representative lines of \ion{Li}i (6707\,\AA, \textit{top left}), \ion{Al}i (3961\,\AA, \textit{top right}), \ion{Ca}{ii} (3933\,\AA, \textit{bottom left}), and \ion{Fe}i (3719\,\AA, \textit{bottom right}) in a high-resolution snapshot of the hydrodynamical model ($240^2$ grid points for Li, Al and Ca; $120^2$ for Fe). Although the behavior is shown for vertical rays, we emphasize that the statistical equilibrium was solved using the 3D radiation field.
The NLTE line strength correction, $\Delta \log W_{\lambda,\mu=1} = \log W_{\lambda,\mu=1} (\text{NLTE}) - \log W_{\lambda,\mu=1} (\text{LTE})$, corresponds roughly to the abundance correction but with opposite sign. Red and blue colors indicate weaker and stronger lines under NLTE than LTE, meaning positive and negative abundance corrections.
For this illustration, we adopt the abundances $\Abund{Li} = 0.70$, $\Abund{Al} = 1.15$, $\Abund{Ca} = -0.62$, and $\Abund{Fe} = 0.60$. Na, not shown, behaves very similarly to Li, and the corresponding illustration for Mg is shown in the bottom left of Fig.~\ref{fig:resolved_EWs}.} 
\label{fig:dNLTE}
\end{figure*}

We illustrate the predicted NLTE line strength corrections at disk-center (\ie using vertical rays, where $\mu = 1$) on the resolved surface of a high-resolution snapshot from the hydrodynamical simulation for representative lines of \ion{Li}i, \ion{Al}i, \ion{Ca}{ii} and \ion{Fe}i in Fig.~\ref{fig:dNLTE}. 
The NLTE line strength correction, $\Delta \log W_{\lambda,\mu=1} = \log W_{\lambda,\mu=1} (\text{NLTE}) - \log W_{\lambda,\mu=1} (\text{LTE})$, represents the difference in line strength between the NLTE and the LTE case, and is approximately the same as the abundance correction (but with opposite sign) for weak lines. Note however that line strengths vary from center to limb, with typically stronger NLTE--LTE differences toward the limb as rays with small $\mu$ probe higher atmospheric layers where collisional rates are lower.

Below, we give detailed descriptions on the NLTE behavior of each species at disk-center.
We find the NLTE corrections across the surface of the simulation to be negative across the bulk of the hot granules. The overall behavior varies between elements due to different processes being responsible for NLTE effects, with our different diagnostic lines forming from levels of different excitation energy as well as ionization stages.

\subsubsection{Lithium and sodium}
At low abundances, the Li and Na resonance lines behave similarly as they form in the transition region between over-ionization and over-recombination. 
Over-ionization tends to dominate in granules while over-recombination gives a net line strengthening in the intergranular lanes. 
The interplay depends on the line strength, as the ground state tends to be over-populated in deeper layers but depleted in higher layers. Since stronger lines form in higher layers, the NLTE correction becomes more negative the higher the abundance. As long as the lines are very weak, this ionization balance effect is more important than photon losses in the core.

\subsubsection{Magnesium} \label{sect:mg}
\begin{figure*}
\centerline{\includegraphics[clip,trim=1em 1em 0.3em 1em]{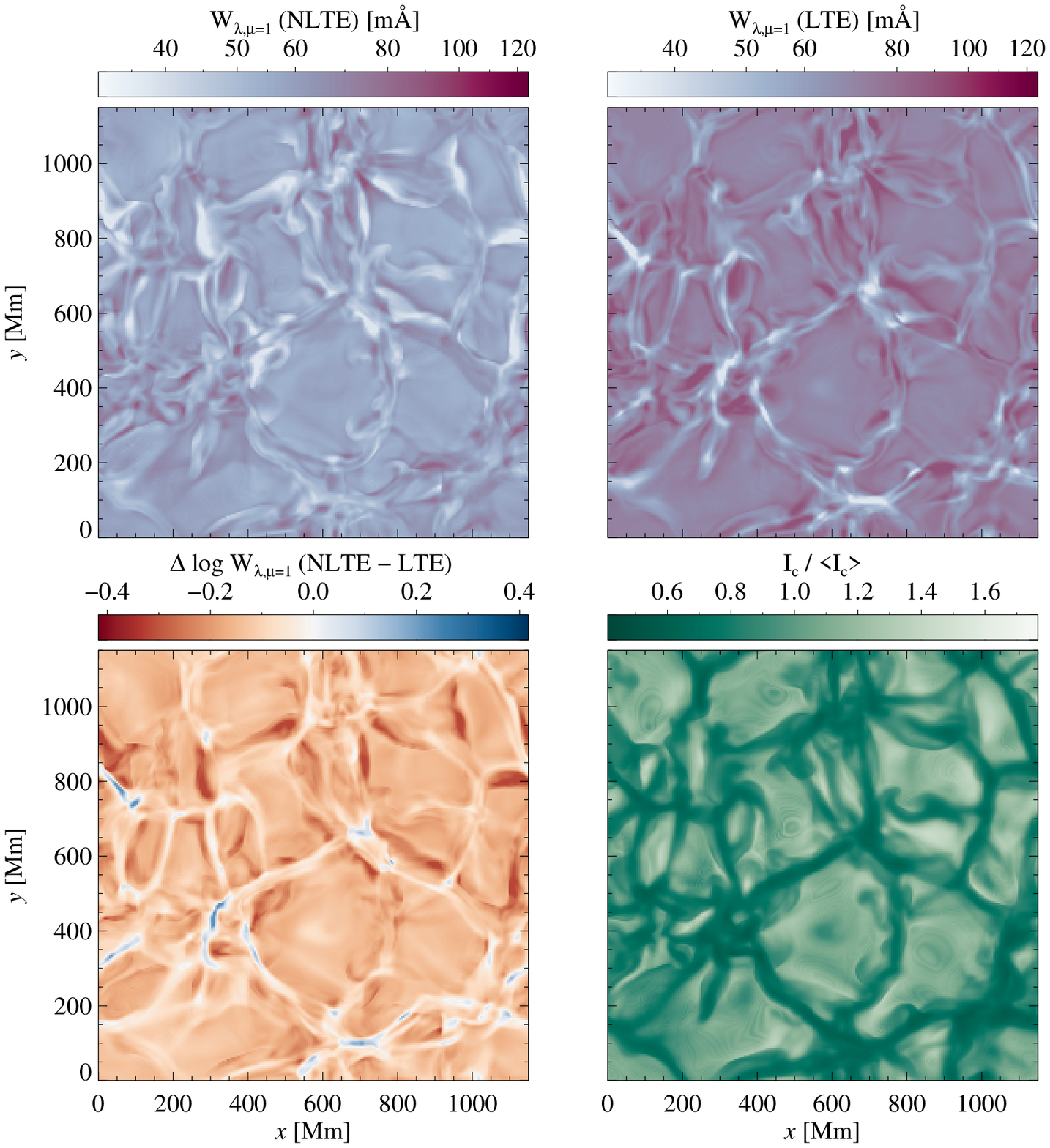}}
\caption{Spatially resolved line equivalent widths of the \ion{Mg}{i} 5183\,\AA\, line at disk-center intensity ($\mu = 1$) in a high-resolution ($240^2$) snapshot of the hydrodynamical model, assuming $\Abund{Mg} = 3.80$.
\textit{Top left}: NLTE (\textit{right:} LTE) equivalent widths. 
\textit{Bottom left}: NLTE equivalent width correction. Red and blue colors carry the same meaning as in Fig.~\ref{fig:dNLTE}. 
\textit{Bottom right}: Granulation pattern imprinted on the relative continuum intensity.
} \label{fig:resolved_EWs}
\end{figure*}

We illustrate the predicted line strength for the \ion{Mg}ib 5183\,\AA\ line on the resolved surface of the hydrodynamical simulation in Fig.~\ref{fig:resolved_EWs}. 
In LTE, the predicted line strength correlates with the continuum intensity $I_\text{c}$, \ie the optical granulation pattern, such that the line is typically stronger in the bright, granular upflows than in the cooler intergranular lanes. This is because the saturated line core forms high above the continuum, where temperature inversions above intergranular lanes reduce line strengths. The NLTE line strength in the granules is however clearly suppressed due to over-excitation driven mainly by the UV triplet at 3829--3838\,\AA, which efficiently depopulates the \conf{3p}{^3P^o} states. The fact that the NLTE effect is strongest in the hot regions is expected, as over-excitation and over-ionization result from the $J_\nu-B_\nu$ split, which is larger where the temperature structure is steeper. Consequently, this effect is typically smaller in the intergranular lanes, with cooler regions even experiencing enhanced line strength in patches with significant temperature inversions.

\subsubsection{Aluminium}
The ground state of Al is efficiently de-populated relative to LTE by over-ionization and over-excitation driven by the resonance lines.
The NLTE line strength correction varies strongly between essentially zero in the cool intergranular lanes and as much as $-0.7$\,dex in the granules, where the strong temperature gradient results in a large $J_\nu-B_\nu$, which drives the NLTE effect.

\subsubsection{Calcium}
For Ca, the dominant NLTE driver is over-excitation of the upper state, \conf{4p}{^2P^o}, driven by the resonance lines. This has minute effects on the ground state but strongly over-populates the upper states. A flow back down via the IR triplet lines over-populates the \conf{3d}{^2D} state as well, which is collisionally but not radiatively coupled to the ground state. 
The resonance line formation behaves like a two-level atom, with a superthermal source function (due to the population inversion) equal to the mean intensity $\bar J_\nu$, resulting in weakened lines. 
Aside from the same type of saturation effect as was seen for Mg in Fig.~\ref{fig:resolved_EWs} in regions where the temperature structure is inverted in layers high above the continuum, NLTE effects on \ion{Ca}{ii} vary only weakly across the surface.

\subsubsection{Iron}
NLTE effects on the Fe lines are similar to those seen for Al. The same drivers, photon pumping in the resonance lines and over-ionization, are responsible for the departures from LTE. However, while photoionization operates on the ground state of \ion{Al}i, it is mainly the excited states of \ion{Fe}i which are over-ionized.
NLTE effects on the \ion{Fe}i resonance lines are thus due to the resonance lines driving a flow from the ground state to the intermediate states, which are over-ionized.
The NLTE line strength correction is stronger (more negative) than for Al, with a broad distribution of line strength corrections peaking at $-1$\,dex but ranging to $-1.5$\,dex.

\subsection{3D NLTE effects on disk-integrated flux spectra} \label{sect:flux}
Line strengths computed in NLTE and LTE using the highest resolution ($240^2$, except for Fe where $120^2$ was used) snapshot from the hydrodynamical simulation are indicated in Fig.~\ref{fig:strengthvar} by horizontal dashed lines. 
NLTE line strengths are smaller (giving positive abundance corrections), with differences ranging from just 0.04\,dex for the \ion{Na}i 5895\,\AA\ line to 1.15\,dex for \ion{Fe}i 3719\,\AA. 
We caution the reader that the NLTE effects vary between lines of the same species, as well as with abundance, and cannot necessarily be directly applied to an abundance analysis.
Additionally, the lines of Mg and Ca are saturated, leading to larger effects on abundances than what is indicated by differences in line strengths.

In the following subsections, we compare these 3D NLTE results to those computed with 1D or \avgtd\ models, to calculations using model atmospheres reduced in size as well as other snapshots from the hydrodynamical simulation, and evaluate the influence of non-vertical radiative transfer in the NLTE solution.

\newcommand\lcut{2.2} 
\newcommand\rcut{1} \newcommand\rtrim{1} 
\newcommand\bcut{2.3} 
\begin{figure*}[!t]
\centerline{
	\includegraphics[clip,trim=0 \bcut em \rcut em 0]{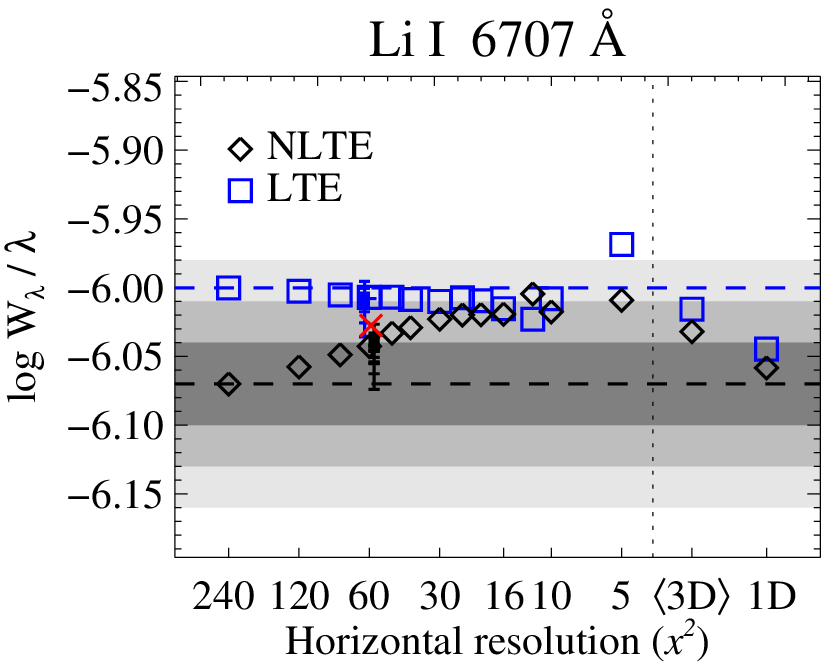}
	\includegraphics[clip,trim=\lcut em \bcut em \rtrim em 0]{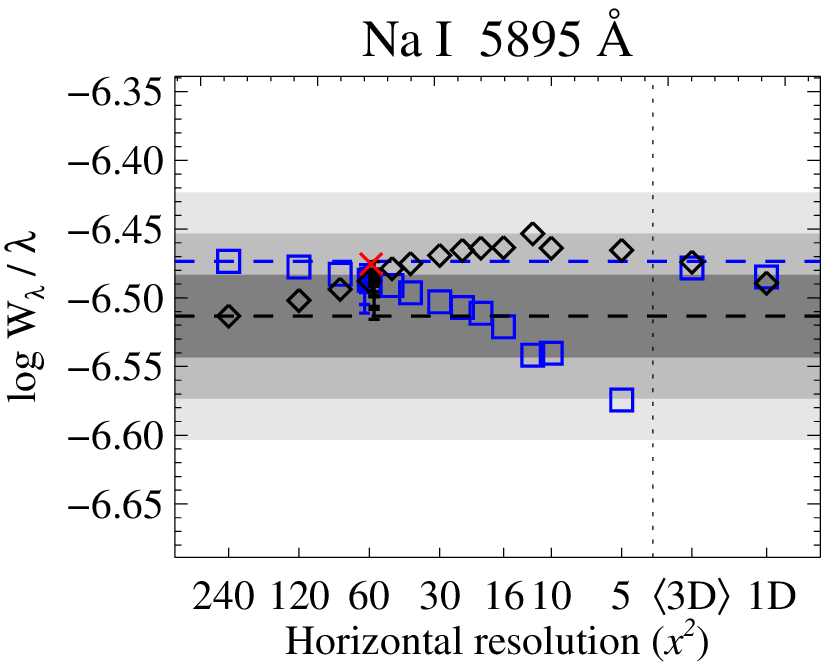}
} \vspace{1em}
\centerline{
	\includegraphics[clip,trim=0 \bcut em \rcut em 0]{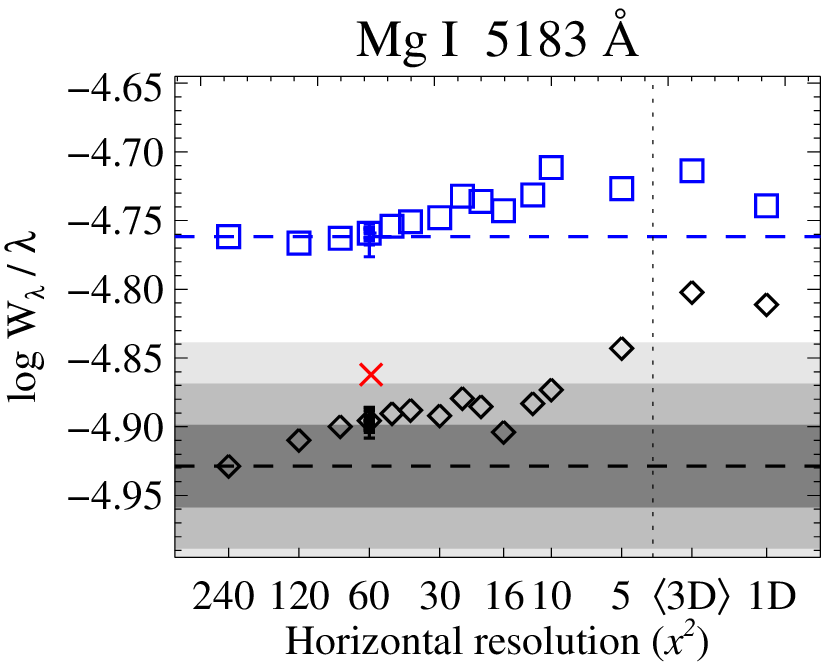}
	\includegraphics[clip,trim=\lcut em \bcut em \rtrim em 0]{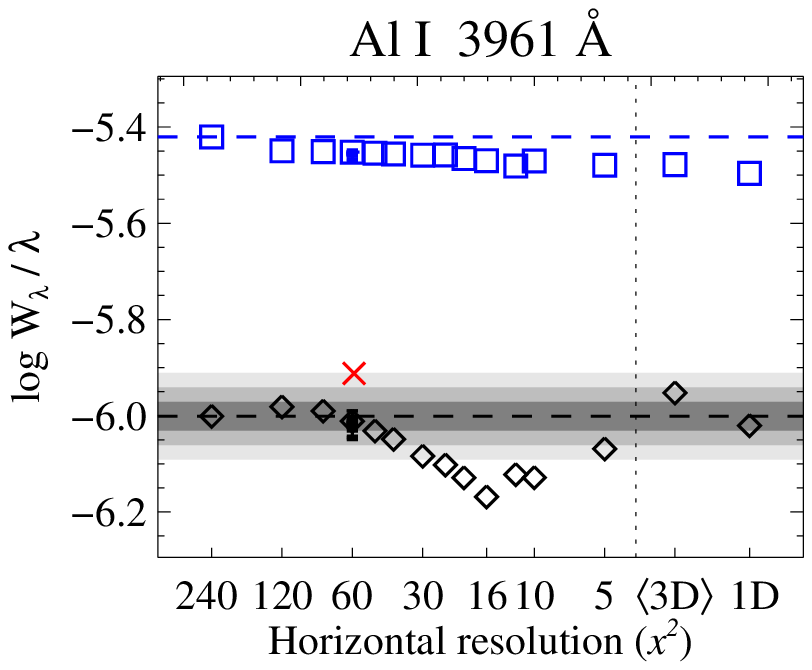}
} \vspace{1em}
\centerline{
	\includegraphics[clip,trim=0 0 \rcut em 0]{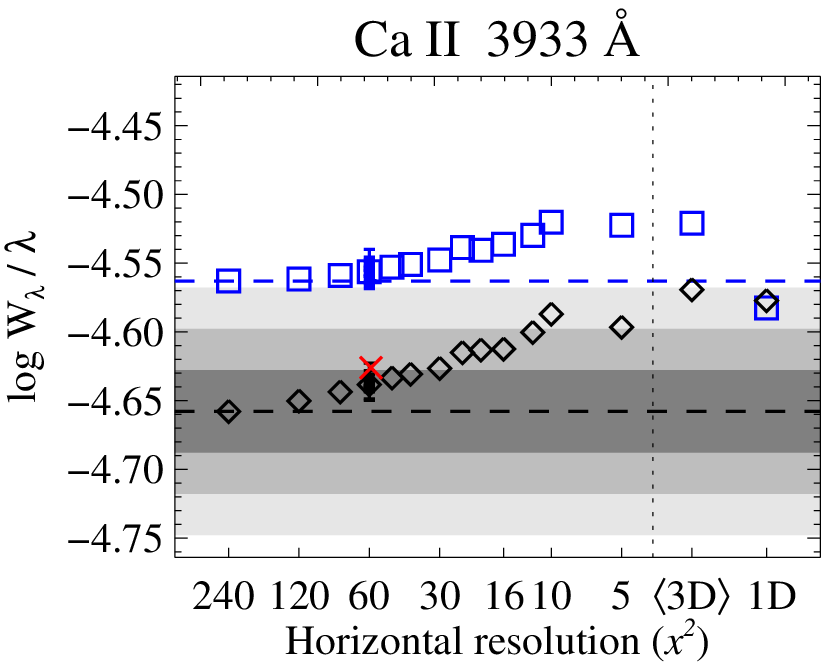}
	\includegraphics[clip,trim=\lcut em 0 \rtrim em 0]{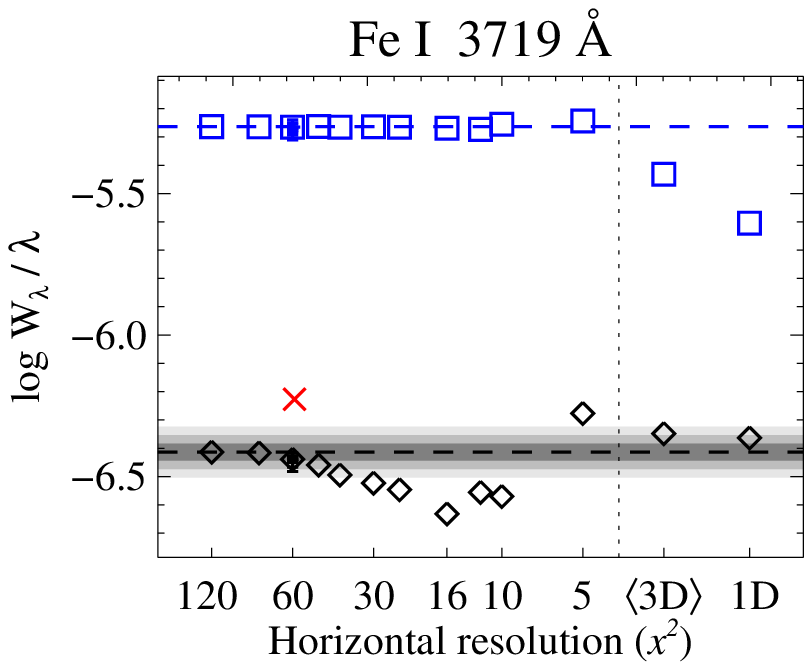}
}
\caption{Variation of predicted disk-integrated line strengths, as inferred from a snapshot from the hydrodynamical simulation downsampled from the original $240^2$ grid points as well as the horizontally averaged \avgtd\ and hydrostatic 1D case. 
A vertical dashed line separates the 3D results from the 1D hydrostatic and \avgtd\ results. Calculations in NLTE (black diamonds) and LTE (blue squares) are illustrated for representative lines using $\Abund{Li} = 0.70$, $\Abund{Na} = 0.89$, $\Abund{Mg} = 3.80$, $\Abund{Al} = 1.15$, $\Abund{Ca} = -0.62$, and $\Abund{Fe} = 0.60$. 
When additional snapshots have been computed (at resolution $60^2$), their individual results are indicated by horizontal lines.
A red greek cross indicates the corresponding NLTE result computed in the so-called 1.5D approximation (at resolution $60^2$).
Each panel spans over 0.35\,dex, except for Al and Fe due to their very strong NLTE effects. 
For clarity, the line strength determined from the highest-resolution model is indicated as a horizontal dashed line across each panel, and representative errors of 0.03, 0.06 and 0.09\,dex are indicated by the gray shaded regions. }
\label{fig:strengthvar}
\end{figure*}

\subsubsection{3D--1D comparison}
Line strengths computed using \avgtd\ horizontally and temporally averaged one-dimensional structure, and the 1D hydrostatic MARCS model, are shown in Fig.~\ref{fig:strengthvar}.
Comparing LTE results, we find that \avgtd\ models reproduce the 3D LTE results well, with deviations of less than 0.05\,dex except for Fe where lines are too weak by 0.2\,dex. 
In \avgtd\ NLTE, lines of Li, Na, Al and Fe are too strong by 0.04--0.07\,dex compared to 3D NLTE. The Ca line is too strong by 0.11 dex, and the lines of Mg are too strong by 0.12--0.17\,dex depending on line. 

For MARCS models in LTE, differences with respect to 3D LTE are larger than for \avgtd\ models, as expected, with Fe line strengths being 0.3\,dex too weak.
In NLTE, results for MARCS models are very close to 3D NLTE for Li, Na and Al, to within 0.02\,dex. Lines are too strong by 0.05\,dex for Fe, 0.10\,dex for Ca, and 0.10--0.16\,dex for Mg. 

Overall, NLTE predictions for weak lines using these 1D MARCS or \avgtd\ models are not too different from the full 3D NLTE solution. 
For the saturated lines of Mg and Ca however, line strengths are significantly over-estimated, indicating that the adopted value $\eqnvmic = 2$\,\kms, may be too high. As the curve of growth is flat for saturated lines, the resulting effect on abundances is even larger than that indicated by the difference in line strength. We discuss the choice of \vmic\ parameter further in Sect.~\ref{sect:abundances}.

\subsubsection{3D modeling errors} \label{sect:numerics}
We investigate both the influence of reducing the resolution of a given snapshot from the hydrodynamical simulation, and the temporal snapshot-to-snapshot variations, using Fig.~\ref{fig:strengthvar}.

Reducing the resolution appears to introduce systematic errors in both NLTE and LTE calculations.
NLTE results for lower-resolution models systematically overestimate line strengths for Li, Na, Mg and Ca, while underestimating them for Al and Fe. 
The magnitude of these effects appears larger for lines where NLTE effects are large. Errors are smaller in LTE, and are similar for all lines.

We find that for 3D NLTE modeling, the model atmospheres can safely be reduced in resolution to $60^2$, with errors less than 0.03\,dex compared to calculations at the maximum resolution.
For \ion{Fe}i, we have not been able to solve the statistical equilibrium for models at higher resolution than $120^2$. 
Line strengths deduced from these models differ from results at resolution $60^2$ by only 0.02\,dex. 
As the NLTE effects are similar to those seen for \ion{Al}i, albeit stronger, we estimate that the errors introduced by running these simulations at lower resolution should not be larger than $\sim$0.03\,dex. 

Little work has been done to test the sensitivity on abundance analyses to the numerical resolution used in the 3D stellar models and in the radiative transfer calculations.
\citet{asplund_effects_2000} studied iron lines in the solar spectrum, where the low-resolution models ($50^2$ compared to $200^2$) were 50--100\,K warmer in the outer atmospheric layers. In their LTE synthesis, this resulted in weaker lines indicating higher abundances by $\sim$0.05\,dex, but reaching as high as 0.1\,dex for low-excitation lines.
Systematic temperature differences of similar magnitude have been found at higher resolution ($140^2 \times 150$ compared to $400^2 \times 300$) in simulations of the solar photosphere computed with the \cobold\ code \citep{freytag_simulations_2012}.
Based on the small effects seen at very high resolution ($480^2$) in models computed for HD~122563 by \citet{collet_abundance_2009}, we expect that our hydrodynamical model is sufficiently large in this respect.

In addition to the numerical resolution, we have analyzed sequences of snapshots to estimate the effect of temporal variations, and indicate the line strengths inferred from individual snapshots by horizontal lines in Fig.~\ref{fig:strengthvar}. 
The snapshot-to-snapshot RMS variation at $60^2$ resolution is less than 0.02\,dex.
The set of five snapshots we use for the abundance analysis does sample the simulation well, and even in the case of \ion{Fe}i where we use a single snapshot, the small sample of snapshots does not introduce significant errors.

In summary, we find that it is unlikely that combined errors due to the limited resolution and number of snapshots is larger than 0.05\,dex for any species analyzed here, as compared to the full $240^2$ model. 
We find smaller errors for \ion{Li}i, \ion{Na}i, \ion{Mg}i and \ion{Ca}{ii}, with combined effects likely not larger than 0.02\,dex.
The 3D and NLTE effects are thus much greater than these numerical errors.
Additionally, these errors fall well within the estimated error bars associated with uncertainties in the stellar parameters, indicated in Table~\ref{tbl:sensitivity}.

Other potential sources of errors include \eg, effects of the opacity binning and NLTE effects on hydrogen which may affect the atmospheric structure, and input data such as photoionization and collisional transition rates in the NLTE modeling.
Rigorous propagations of atomic data uncertainties are beyond the scope of this analysis, but see \eg \citet{osorio2015} where the influence of different collisional processes was explicitly tested for Mg.

\subsubsection{The influence of non-vertical radiative transfer}
To help gauge the influence of non-vertical radiative transfer on the NLTE solution, we have also performed test calculations using the so-called 1.5D approximation, where the NLTE solution is acquired separately column-by-column in the model atmosphere. 
The 1.5D approximation can also be utilized to save computational resources, as a full 3D solution of the statistical equilibrium is avoided \citep[\eg][]{kiselman_777_1993,shchukina_iron_2001,shchukina_impact_2005,shchukina_solar_2009,pereira_rh_2015}.
In the case of the 777\,nm oxygen triplet lines, \citet{amarsi2016} found that solving the statistical equilibrium using the full 3D radiation field did not differ greatly from the 1.5D approximation, but this is not a general result applicable to any atom or type of star.

We indicate in Fig.~\ref{fig:strengthvar} line strengths computed in the 1.5D approximation, and find that two main mechanisms appear to be important. 
With the full 3D radiation field, regions above the cool intergranular lanes are illuminated by the nearby hot granules via inclined rays. As these rays are missing in 1.5D, the $J_\nu - B_\nu$ split decreases, thus leading to strengthened lines in these regions. 
Conversely, the radiation field above the hot granules is somewhat stronger in 1.5D, leading to weakened lines in these regions. 
We find that the net result is a slightly more positive NLTE line strength correction (leading to lower abundances) for all six elements, albeit with effects of at most 0.02\,dex for Li, Na and Ca, 0.04\,dex for Mg, 0.10\,dex for Al, and 0.22\,dex for Fe.

\subsection{Abundance analysis} \label{sect:abundances}

\begin{table*}
\caption{Results of the 3D NLTE abundance analysis.}
\label{tbl:results}
\centering
\begin{tabular}{l r@{}lr@{}l r@{}lr@{}l r@{}lr@{}l r@{}lr@{}l}
\hline\hline\noalign{\smallskip}
Element      & \multicolumn2c{3D NLTE}&\multicolumn2c{3D LTE}& \multicolumn2c{\avgtd\ NLTE} & \multicolumn2c{\avgtd\ LTE} & \multicolumn2c{1D NLTE} & \multicolumn2c{1D LTE} & \multicolumn2c{\citet{bessell2015}}  \\
\hline \noalign{\smallskip}
\Abund{Li} \tablefootmark a & $ 0.82$&$\pm0.08$ & $ 0.79$&$\pm0.08$ & $ 0.81$&$\pm0.07$ & $ 0.79$&$\pm0.07$ & $ 0.83$&$\pm0.07$ & $ 0.82$&$\pm0.07$ & $0.71$&$\pm0.10$ \\
$\Abund{Na}$ \tablefootmark b &$< 0.87$ & \, ($ 0.64$)&$< 0.87$&          &$< 0.85$&          &$< 0.86$&          &$< 0.87$&          &$< 0.86$&       & $< 0.58$&          \\
\Abund{Mg} \tablefootmark c & $ 3.77$&$\pm0.03$ & $ 3.40$&$\pm0.06$ & $ 3.56$&$\pm0.08$ & $ 3.28$&$\pm0.12$ & $ 3.59$&$\pm0.07$ & $ 3.28$&$\pm0.11$ & $3.52$&$\pm0.03$ \\
$\Abund{Al}$ \tablefootmark b &$< 1.30$ & \, ($ 0.79$)&$< 0.69$&          &$< 1.22$&          &$< 0.72$&          &$< 1.29$&          &$< 0.74$&     &$<0.25$           \\
\Abund{Ca} \tablefootmark a & $-0.60$&$\pm0.06$ & $-1.14$&$\pm0.04$ & $-0.96$&$\pm0.05$ & $-1.28$&$\pm0.03$ & $-0.91$&$\pm0.04$ & $-1.04$&$\pm0.05$ &$-0.92$&$\pm0.10$ \\
$\Abund{Fe}$ \tablefootmark {b,d} &$< 0.97$ &\, ($ 0.50$)&$<-0.18$&          &$< 0.89$&          &$<-0.01$&          &$< 0.77$&          &$< 0.16$&    &$<-0.02$       \\
\hline
\end{tabular}
 \tablefoot{
The recommended abundances from \citet{bessell2015} are shown for comparison. These are based on 3D LTE for Li, \avgtd NLTE for Na, Mg, Ca and Fe, and 1D LTE for Al. Note that the upper limits are given as $1 \sigma$ for Na and Al, and $3 \sigma$ for Fe. \\
	 \tablefoottext{a}{Line profile fits, with $1 \sigma$ uncertainties from the $\chi^2$ statistic.}
	 \tablefoottext{b}{$3 \sigma$ upper limits based on the local photon noise and continuum placement uncertainty. The $1 \sigma$ upper limit is shown in parentheses for the 3D NLTE analysis.}
	 \tablefoottext{c}{Average of multiple line profile fits, with $1 \sigma$ line-to-line dispersion.}
	 \tablefoottext{d}{$3 \sigma$ upper limits based on stacked spectra of multiple unblended lines.}
 }
\end{table*}

\begin{table}
\caption{Recommended abundances of \thestar, with statistical and systematic uncertainties}
\label{tbl:abundances}
\centering
\begin{tabular}{l r r l c c}
\hline\hline \noalign{\smallskip}
Species      & \Abund X & \eH X & $\sigma_\text{stat}$ & $\sigma_\text{sys}$\tablefootmark{a} & Source \\
\hline \noalign{\smallskip}
\ion{Li}{i}& $ 0.82$ & $-0.23$& 0.08 & 0.10 & * \\
C (CH)& $ 5.88$ & $-2.55$& 0.09 && 1 \\
N (NH)&$< 3.13$  &$<-4.70$& 0.3   && 1 \\
O (OH)& $ 6.16$& $-2.53$& 0.15 && 1 \\
\ion{Na}{i}&$< 0.64$  &$<-5.60$ &  0.10 & 0.08   & * \\
\ion{Mg}{i}& $ 3.77$ & $-3.83$& 0.03 & 0.10   & * \\
\ion{Al}{i}&$< 0.79$  &$<-5.66$& 0.25 & 0.11  & * \\
\ion{Si}{i}&$< 1.91$  &$<-5.60$&  0.3  && 2 \\
\ion{Ca}{ii}& $-0.60$ & $-6.94$& 0.06  & 0.10  & * \\
\ion{Sc}{ii}&$<-2.35$  &$<-5.50$&  0.3  && 2 \\
\ion{Ti}{ii}&$<-2.15$  &$<-7.10$&  0.3  && 2 \\
\ion{V }{ii}&$<-0.37$  &$<-4.30$&  0.3  && 2 \\
\ion{Cr}{i}&$<-0.66$ &$<-6.30$&   0.3 && 2 \\
\ion{Mn}{i}&$<-0.37$  &$<-5.80$&  0.3  && 2 \\
\ion{Fe}{i}\tablefootmark b &$< 0.50$  &$<-7.00$&  0.24  & 0.16  & * \\
\ion{Co}{i}&$<-0.41$  &$<-5.40$&   0.3 && 2 \\
\ion{Ni}{i}&$<-0.68$  &$<-6.90$&  0.3  && 2 \\
\ion{Cu}{i}&$<-1.31$   &$<-5.50$&  0.3  && 2 \\
\ion{Zn}{i}&$< 1.16$  &$<-3.40$&   0.3 && 2 \\
\ion{Sr}{ii}&$<-4.03$  &$<-6.90$&  0.3  && 2 \\
\ion{Ba}{ii}&$<-3.92$  &$<-6.10$&   0.3 && 2 \\
\ion{Eu}{ii}&$<-2.38$  &$<-2.90$&   0.3 && 2 \\
\hline
\end{tabular}
 \tablefoot{
Uncertainties and upper limits are given at the $1 \sigma$ level.
	\tablefoottext{a}{Systematic uncertainty due to uncertainties in stellar parameters, estimated by \citet{keller2014} as $\sigma(\eqnTeff) = 100$\,K and $\sigma(\eqnlogg) = 0.2$\,dex.}
	\tablefoottext{b}{The corresponding $3 \sigma$ upper limit is $\Abund{Fe} < 0.97$, $\eqnFeH < -6.53$.}
 }
 \tablebib{
 (*) This work, 3D NLTE;
 (1) \citet{bessell2015}, 3D LTE analysis of molecular lines;
 (2) \citet{bessell2015}, 1D LTE analysis.
 }
\end{table}

\begin{figure}
	\centerline{\includegraphics[width=8.8cm]{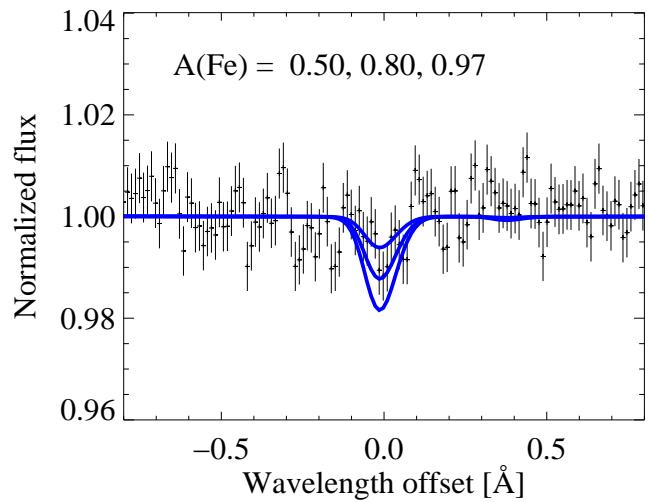}}
	\caption{Upper limit determination for iron, comparing stacked spectra centered on the six strongest unblended lines, \ion{Fe}i 3440.6, 3581.2, 3719.9, 3737.1, 3820.4 and 3859.9\,\AA. The 3D NLTE line profiles are shown for the 1$\sigma$, 2$\sigma$ and 3$\sigma$ upper limits, corresponding to line strengths of 0.7, 1.5 and 2.2\,m\ang. The small bump at $+0.4\,\AA$ in the synthetic spectrum is due to the nearby line \ion{Fe}i 3441.0\,\AA.}
	\label{fig:fe}
\end{figure}

The results of our 3D NLTE analysis are shown in Table~\ref{tbl:results}, and illustrated for iron in Fig.~\ref{fig:fe}. The stacked spectrum of the \ion{Fe}i 3440.6, 3581.2, 3719.9, 3737.1, 3820.4 and 3859.9\,\AA\ lines does hint at a weak absorption feature, but we stress that it is consistent with the noise level at $1\sigma$.
For comparison, we also report in Table~\ref{tbl:results} the corresponding LTE results, as well as results from one-dimensional modeling using \avgtd and 1D MARCS models.
We combine our results with abundances of C, N and O based on 3D LTE molecular line analyses, and 1D LTE analyses for the remaining elements from \citet{bessell2015} in Table~\ref{tbl:abundances}. These are our recommended abundances for comparisons to supernova models.

Differences between our 3D NLTE results and abundances determined in LTE using 1D MARCS models are negligible for Li and Na, but amount to $+0.5$\,dex for Mg, Al and Ca, and $+0.8$\,dex for Fe. 
Comparing instead to 1D NLTE results, we find excellent agreement for Li, Na and Al, and differences decrease somewhat to $+0.2$\,dex for Mg and Fe, and $+0.3$\,dex for Ca.
In contrast, abundances based on the 3D LTE analysis deviate more strongly from 3D NLTE than did the 1D LTE analysis, indicating that 3D LTE analyses of atomic lines are unsuitable in the presence of strong NLTE effects. 
This is expected, as the steeper temperature structures of the 3D models give rise to even stronger NLTE effects (as is evident from Table~\ref{tbl:results} for every element except Na).

Abundance analyses in LTE using the temporally and horizontally averaged \avgtd\ model, which has a steeper temperature gradient than the 1D MARCS models, again deviate more strongly from the 3D NLTE results than did the 1D LTE analysis, and are more similar to 3D LTE. 
In NLTE, results with the \avgtd\ model are similar to the 1D NLTE analysis, indicating that adopting the more realistic temperature structure of the \avgtd\ model is not sufficient, but in fact horizontal inhomogeneities, granulation, must be taken into account to correctly model the NLTE line formation. 
Additionally, results for \avgtd\ models are sensitive to the averaging method \citep{magic2013_avg}.

\begin{figure}
	\centerline{\includegraphics[width=8.8cm]{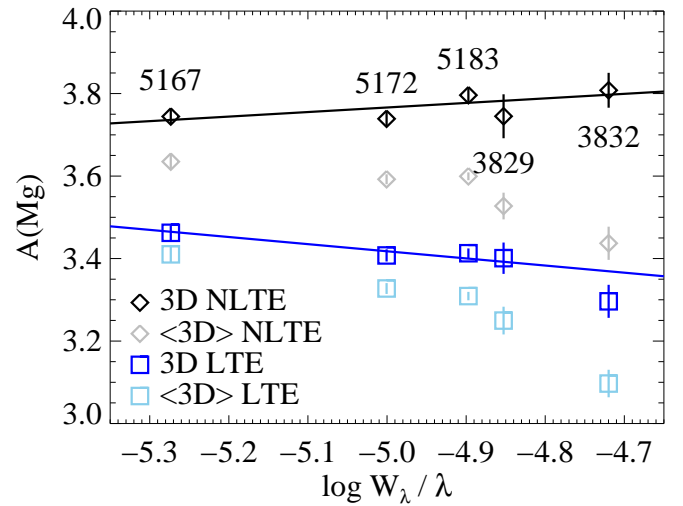}}
	\caption{Abundance results for five lines of magnesium using different types of modeling, shown as a function of the reduced equivalent width. Line central wavelengths are indicated in units of \AA. For the 3D NLTE and 3D LTE cases, we overplot linear fits to the trends. In 3D NLTE, we find a positive slope of less than 2$\sigma$ significance, while 3D LTE as well as the \avgtd\ results indicate $>$3$\sigma$ significant negative slopes.}
	\label{fig:mgabund}
\end{figure}

As illustrated in Fig.~\ref{fig:mgabund}, Mg abundances derived from the individual lines in 3D NLTE are in excellent agreement, with a scatter of only 0.03\,dex, weakly correlated with line strength. 
3D LTE results, on the other hand, indicate significantly lower abundances derived from the stronger lines. \avgtd\ NLTE and LTE abundances also strongly anticorrelate with line strength; we find very similar behavior using 1D MARCS models. 
As all lines but the weakest are saturated, 1D and \avgtd results are sensitive to the choice of the microturbulence parameter, which we set to $\eqnvmic = 2$\,\kms. 
Decreasing \vmic\ by 0.5\,\kms\ would increase the abundance determined from the strongest line by 0.10\,dex.
Thus, in order to bring lines of \ion{Mg}i of different strength into agreement, \vmic\ would need to decrease significantly, to 1\,\kms.
Such a low value would increase the average deduced Mg abundance by 0.1\,dex, bringing \avgtd\ NLTE results into agreement with 3D NLTE. 
Setting $\eqnvmic = 1$\,\kms\ would however also weaken the heavily saturated \ion{Ca}{ii} 3933\,\AA\ line by 0.1\,dex, causing an enormous 0.6\,dex increase in the inferred abundance.
While we do not use the \ion{Ca}{ii} 3968\,\AA\ line in our abundance analysis, we note that results from the two lines agree to within 0.01\,dex in 3D NLTE. Setting $\eqnvmic = 1$\,\kms\ would lead to a 0.2\,dex disagreement between the two \ion{Ca}{ii} lines in \avgtd\ NLTE. 
Therefore, the free parameter \vmic\ cannot be invoked to simultaneously flatten trends with line strength of both \ion{Mg}i and \ion{Ca}{ii}, nor to bring 3D NLTE and \avgtd\ NLTE synthesis of these saturated lines into agreement.

Compared to the results of \citet{keller2014} and \citet{bessell2015}, our abundances are typically slightly higher when running comparable analysis methods. 
To explain this, we first note that our tailored atmospheric models were computed using a chemical composition similar to \thestar, affecting the overall structure, with somewhat higher \Teff\ (by 25\,K) and lower \logg\ (by 0.1\,dex) than was determined by \citet{keller2014}. 
Our abundance analysis is based on spectra computed with \multitd using the hydrodynamical model. In contrast, \citet{keller2014} and \citet{bessell2015} used 1D models interpolated to the adopted stellar parameters in their 1D LTE analyses, and then applied representative abundance corrections based on comparisons between 1D LTE and 3D LTE or \avgtd\ NLTE modeling. 
While differences in stellar parameters are well within the estimated observational error bars of 100\,K and 0.2\,dex, they lead to us determining higher abundances by 0.02--0.04\,dex, or lower by 0.05\,dex for Ca, based on the sensitivities listed in Table~\ref{tbl:sensitivity}.
Second, the UVES spectra were aligned and coadded anew. 
Upper limits to the faint features of Al, Na and Fe are determined using Cayrel's formula \citep{Cayrel1988,Cayrel2004} taking into account uncertainties in the continuum placement, and for Fe we stack a larger number of spectral lines.

Our \avgtd NLTE abundances of Li, Na, Mg and Al agree to within 0.1\,dex with \citet{bessell2015}. Discrepancies are likely due to differences in the data reduction, continuum normalization, and in the case of Na and Al the upper limit determinations. For Mg, our 3D NLTE abundance is higher than the previous best estimate by 0.25\,dex.
For Al, our 3D NLTE abundance limit ($1 \sigma$) is higher by 0.5\,dex than the previous 1D LTE value of \citet{bessell2015}. This is entirely due to the 3D NLTE effect, as our 1D LTE $1 \sigma$ upper limits agree to within 0.02\,dex.
Our \avgtd NLTE Ca abundance agrees with previous estimates, and our 3D NLTE abundance is higher than this by 0.3\,dex.

Finally, our $3\sigma$ upper limit for Fe is $\Abund{Fe} = 0.97$ in 3D NLTE, compared to $\Abund{Fe} = -0.02$ as determined in \avgtd NLTE by \citet{bessell2015}.
Our higher abundance limit is mainly due to our more conservative statistical approach to estimating the upper limit ($+0.5$\,dex), new NLTE data ($+0.2$\,dex) and the use of 3D NLTE rather than \avgtd NLTE ($+0.1$\,dex), all of which lead to a higher abundance. This does not however fully explain our higher abundance limit, indicating that additional effects due to differences in \eg synthesis codes, averaging methodology and continuous opacities may be present. Tracing and quantifying these differences is challenging, but we note that our analysis is directly based on 3D NLTE synthesis, rather than 1D LTE synthesis with a posteriori corrections. 
We stress that at $\Abund{Fe} = 0.97$, the stacked 3D NLTE spectrum has an equivalent width of just 2.2\,m\ang, and that reducing the abundance by 0.5\,dex would result in a line strength of 0.7\,m\ang.



\section{Discussion} \label{sect:discussion}

\begin{figure}
	\includegraphics[]{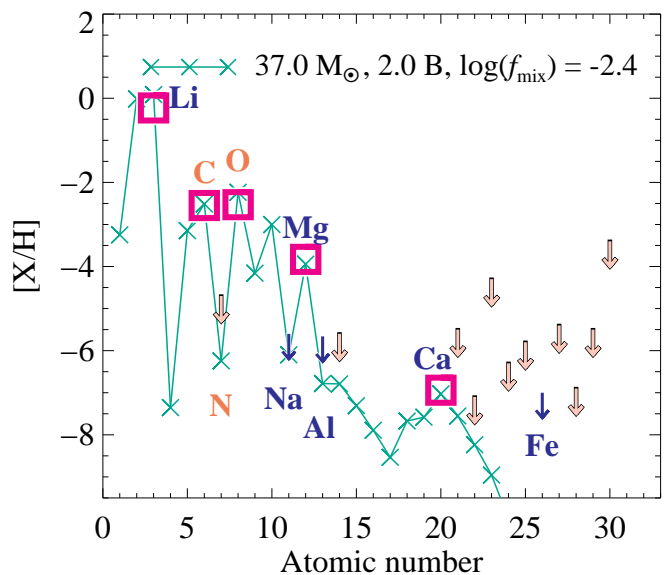}
	\caption{Best fitting core-collapse supernova model yields \citep{heger2010} mixed with primordial BBN-composition gas (see text), compared to the inferred abundances and upper limits listed in Table~\ref{tbl:abundances}. Labeled elements are based on 3D NLTE analysis of atomic lines (blue characters, this work) or 3D LTE analysis of molecular lines \citep[orange characters, from][]{bessell2015}. Other upper limits are based on 1D LTE analyses \citep[from][]{bessell2015}. 
The progenitor stellar mass, explosion energy, and amount of mixing of the best-fit model are indicated in the legend.}
	\label{fig:abund}
\end{figure}

We compare our recommended abundances (Table~\ref{tbl:abundances}) to core-collapse supernova yields computed for Population III stars \citep[][and subsequent online updates in 2012\footnote{Data and routines are available at \url{http://2sn.org/starfit}}]{heger2010} in Fig.~\ref{fig:abund}. 
The recommended abundances are based on our 3D NLTE synthesis for Li, Na, Mg, Al, Ca and Fe. We use literature data from \citet{bessell2015}, where the abundances of C, N and O are based on 3D LTE synthesis of molecular features. 
Other elements, including most heavier elements ($Z > 20$), have only upper limits derived from 1D LTE analyses. As these upper limits are significantly higher than the supernova yields, we do not expect errors due to unknown 3D NLTE effects on these elements to significantly bias our results.

We consider a scenario where the supernova ejecta enrich primordial gas containing He and Li produced in Big Bang Nucleosynthesis \citep[abundances from ][\ie, $\Abund{Li} = 2.67 \pm 0.06$ and $Y_\text p = 0.247$]{cyburt_big_2016}. The final lithium abundance is then reduced according to the typical difference between observed abundances of Li in metal-poor low-mass stars on the lower RGB, $\Abund{Li} = 1.1$, \citep[\eg,][]{lind_signatures_2009} and the BBN level, a difference of 1.6\,dex.
As it is not possible to accurately predict the amount of intrinsic Li depletion in extremely metal-poor stars \citep[see, \eg,][]{sbordone2010,melendez_observational_2010}, we do not use Li in the model constraint.
The shown yield for H indicates the logarithmic enrichment factor, \ie the logarithmic ratio of enriched to primordial hydrogen. This represents the dilution of the supernova ejecta, which is a free parameter set to reproduce the observed abundance levels.

We compare yields to observations using the \starfit code \citep{heger2010}.
The yields were computed for stellar evolution models of non-rotating metal-free stars spanning a wide range in mass (9.6--100\,$\Msol$). As a robust physical explosion mechanism is not yet known, the models are ``exploded'' by means of a piston action, exploring a range of discrete energies (0.3--10\,B, where $1\,\text B = 10^{51}\,\text{erg}$). Additionally, these one-dimensional models cannot predict the amount of Rayleigh-Taylor induced mixing, and the models are instead mixed by means of an averaging process parametrized as a fraction of mass relative to the He core, $f_\text{mix}$.
As yields vary smoothly as a function of energy, albeit sometimes rapidly, we interpolate the grids to a more densely spaced set of explosion energies before running \starfit. 
Variations as a function of mass are however rather chaotic and cannot be considered continuous \citep[clearly illustrated in Fig.~10 of][]{heger2010}.

\begin{figure*}
	\centerline{\includegraphics[width=\textwidth]{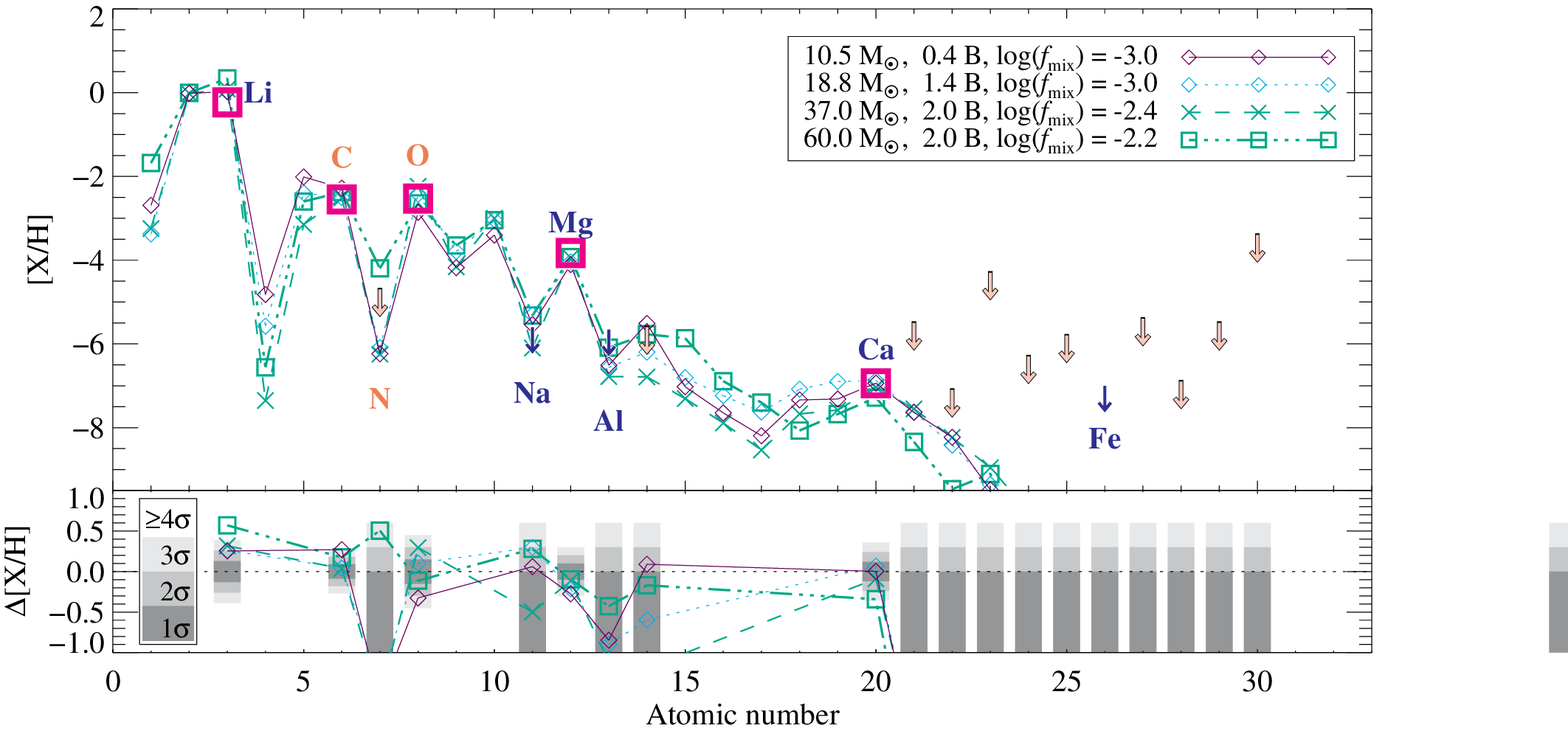}}
	\caption{Best fitting supernova model yields assuming a selection of different progenitor masses, where thicker lines and symbols represent higher mass. The explosion energy (varying color) and mixing fraction (varying plot symbol) have been selected as the best fit to the observed abundances at each selected mass. 
	Residuals are shown beneath for all available elements, with the $1\sigma$, $2\sigma$ and $3\sigma$ measurement uncertainties indicated in dark to light shades of grey.}
	\label{fig:abundvar}
\end{figure*}

The best-fitting supernova model has a progenitor of intermediate mass ($37\,\Msol$), exploding with essentially no mixing ($\log f_\text{mix} \le -2.2$, or 0.6\,\%) and low energy ($2\,$B). 
The low amount of mixing is consistent with predictions from hydrodynamical simulations \citep{joggerst2009}. Increasing the amount of mixing leads to higher yields of Na, Al and Si, and models with $\log f_\text{mix} > -2.2$ produce yields in disagreement with our upper limits.
Due to the low explosion energy, material near the core are not imparted sufficent kinetic energy to leave the gravitational well, but instead fall back into the nascent black hole. The explosion energy thus correlates with yields of heavy elements including O but not C. Increasing the explosion energy by just 10\,\% produces $\eqnXY O C = 0.6$ and $\eqneH{Ca} > -6$, in strong disagreement with observations.

The sensitivity to mass is less straightforward, as we find that a wide range of models can reproduce observations well, but only if the explosion energy is allowed to vary. 
We illustrate the yields from a selection of well-fitting models in Fig.~\ref{fig:abundvar}. The models have been selected as those which best match observations for any explosion energy and mixing efficiency within a range of progenitor masses. 
Hypernovae with explosion energies at 5\,B do not match observations as they over-produce Na and Al while underproducing C and O compared to Mg. At higher energies, hypernovae also strongly overproduce calcium and iron-peak elements. 

More massive stars in the initial mass range 140--260\,$\Msol$ are expected to explode as pair-instability supernovae, with a well-understood explosion mechanism. In the grid by \citet{heger2002}, models show a strong odd-even effect (typical of such supernovae) similar to our observations, but strongly overproduce heavier elements compared to C and O.

\begin{figure}[t]
	\centerline{\includegraphics[]{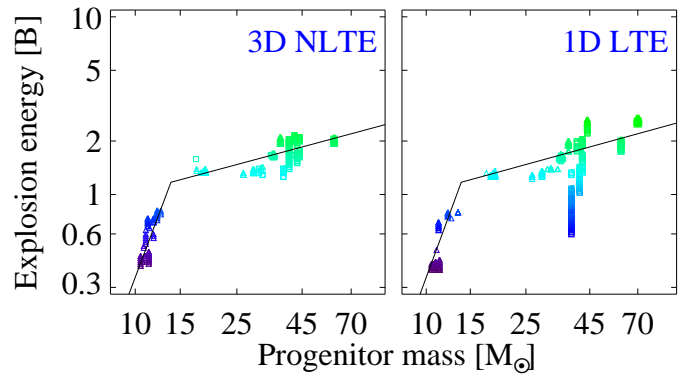}}
	\caption{Comparison of progenitor mass and explosion energy among the well fitted ($\chi^2 < 3$) supernova models. Colors (explosion energy) and symbols (mixing) carry the same meaning as in Fig.~\ref{fig:abundvar}. 
	Results are shown on the \textit{left} for our recommended abundances (from Table~\ref{tbl:abundances}) and on the \textit{right} using 1D LTE abundances. 
	To guide the eye, power-laws fitted to the low-mass and high-mass models in the left panel are shown also to the right.}
	\label{fig:sne}
\end{figure}

The \starfit code selects models using a $\chi^2$ statistic calculated according to eq.~4 of \citet{heger2010}, where we constrain models using 4 measured abundances and 14 upper limits.
We assess that models with fit residuals $\chi^2 < 3$ agree well with observational constraints, resulting in a set of supernova models spanning a wide range of progenitor masses, mixing parameters and explosion energies.
The model parameters are however correlated, and the stronger constraints from our improved spectrum modeling do affect the parameter space of models which reproduce observations. 
We illustrate the co-dependence of parameters in Fig.~\ref{fig:sne}, compared to results from 1D LTE modeling (adopting the same error bars).
Our recommended abundances match a more restricted set of models than the 1D LTE abundances, with better matches to low-mass models. 
The improved modeling is especially restrictive for models with $M \approx 40\,\Msol$, where abundances derived from 1D LTE spectrum modeling match supernova models with explosion energies in the very large range 0.6--2.7\,B. 
We find that low-mass models ($M < 15\,\Msol$) matching observations have experienced essentially no mixing ($\log f_\text{mix} \le -2.2$), and explosion energies are in the range 0.4--0.8\,B, strongly correlated with mass. 
More massive models on the other hand explode with energies in the range 1.3--2.2\,B, while still characterized by little mixing ($\log f_\text{mix} < -1.6$).
We illustrate the relation, and how it compares to 1D LTE abundance results by comparing to fitted power laws with a break near $M = 15\,\Msol$. 

We do not include the abundance of Li in our fit to supernova yields, but note that the model with progenitor mass $60\,\Msol$ shown in Fig.~\ref{fig:abundvar} produces a large overabundance of Li compared to other metals, such that the ejecta mixed with primodial gas is Li-enriched by 0.4\,dex. 
Other models of intermediate mass (40--60\,$\Msol$) and explosion energy (1.4--2.2\,B) produce even more lithium, increasing the abundance by up to 1.6\,dex. 
As Li is created via alpha capture on $^3$He, produced either via muon neutrino spallation of $^4$He or proton captures on neutrons formed in the charged current reaction $\bar \nu_\text e(\text p,\text e^+)\text n$ \citep[see][]{heger2010}, the yield depends on the adopted neutrino fluxes.

\section{Conclusions}
We have analyzed the most iron-deficient star known, \thestar, using 3D NLTE spectrum synthesis for six elements. 
We find generally higher abundances than the previous investigation by \citet{bessell2015}, with a significantly higher upper limit for Fe by 1.0\,dex, now $\eqnFeH < -6.53$ ($3 \sigma$). The increase is mainly due to a more conservative statistical approach ($+0.5$\,dex) and improved NLTE data ($+0.2$\,dex) rather than the use of 3D NLTE synthesis over \avgtd\ NLTE ($+0.1$\,dex), but all these effects happen to increase the inferred abundance.
Our 3D NLTE synthesis also leads to a higher upper limit for Al by 0.5\,dex, now $\eqneH{Al} < -5.66$ ($1 \sigma$), as well as higher abundances for Mg and Ca by 0.3\,dex, now $\eqneH{Mg} = -3.83 \pm 0.10$ and $\eqneH{Ca} = -6.94 \pm 0.12$.

We caution the reader that LTE analyses of extremely metal-poor giant stars, whether in 1D or 3D, are likely to be significantly biased for the elements Mg, Al, Ca and Fe.
These abundances may be underestimated by as much as 1\,dex. 
In particular, our 3D LTE analyses of atomic lines are often \textit{worse} than the 1D LTE analysis, in comparison to the 3D NLTE result, due to cancellation effects in the 1D analysis. 
Such cancellation has previously been predicted \citep[\eg][]{asplund_3d_1999} and later found \citep[\eg][]{asplund_multi-level_2003} in analyses of metal-poor dwarfs, and is similar to the so-called ``NLTE-masking'' demonstrated in a solar 1D analysis by \citet{rutten_empirical_1982}.
Our results from 1D NLTE analyses are similar to those from 3D NLTE, but we find significant differences for the saturated lines of Mg and Ca, which in 1D modeling are sensitive to the \vmic\ parameter, and for Fe where the resonance lines are very sensitive to both NLTE and 3D effects.
Additionally, 3D--1D corrections inferred in LTE and NLTE--LTE corrections inferred in 1D cannot simply be combined, as the NLTE--LTE difference in line strength varies across the surface of the 3D model. 
Unfortunately, NLTE models of neutral species are often limited by the atomic data, in particular transition rates due to collisions with neutral hydrogen atoms. These data are now available for a handful of elements \citep[see][]{barklem_accurate_2016}, the bulk of which we utilize in this work.

Combined with the 3D LTE analysis of molecular lines previously reported by \citet{bessell2015}, our 3D NLTE analysis of atomic lines results in abundances being accurately known for all five elements detected in \thestar (Li, C, O, Mg and Ca) along with accurate upper limits for four more (N, Na, Al and Fe). 
Comparing these abundances to predicted supernova yields \citep{heger2010}, we find good agreement with models of progenitor mass 10\,$\Msol$ and explosion energy $< 1$\,B, or 20--60\,$\Msol$ and explosion energies in the range 1--2\,B, where ejecta from the inner core are unable to escape but fall back into the nascent black hole remnant.
Explosion energies above 2.5\,B can be strongly excluded, as such models eject iron-peak elements producing light-to-heavy element ratios in disagreement with observations.

It is important to extend this work to other ultra metal-poor stars. It is likely that the abundances of other giant stars similar to \thestar are significantly underestimated. The magnitude of the 3D NLTE abundance corrections for iron decrease with increasing iron abundance, such that the abundance scale compresses toward higher values, with \thestar still the most iron-deficient star known by at least an order of magnitude \citep{frebel_near-field_2015}.
Correctly inferring the abundance patterns of ultra metal-poor stars, using 3D NLTE modeling, is crucial to constrain the true nature of Population III supernovae, their progenitor stars, and their influence on the early evolution of galaxies.

\begin{acknowledgements}
We gratefully thank our referee Matthias Steffen for helpful comments which improved the manuscript,
Kjell Eriksson for computing the MARCS model, 
Alexander Heger for discussions on the supernova pollution scenario and details on the supernova yield grid, 
and Conrad Chan for help with the \starfit code.
TN acknowledges support from the Swedish National Space Board (Rymdstyrelsen), and hosting at the ANU for part of this work.
AMA and MA are supported by the Australian Research Council (ARC) grant FL110100012.
KL acknowledges funds from the Alexander von Humboldt Foundation in the framework of the Sofja Kovalevskaja Award endowed by the Federal Ministry of Education and Research as well as funds from the Swedish Research Council (Grant nr. 2015-00415\_3) and Marie Sklodowska Curie Actions (Cofund Project INCA 600398).
PSB acknowledges support from the Royal Swedish Academy of Sciences, the Wenner–Gren Foundation, Goran Gustafssons Stiftelse and the Swedish Research Council. For much of this work PSB was a Royal Swedish Academy of Sciences Research Fellow supported by a grant from the Knut and Alice Wallenberg Foundation. PSB is presently partially supported by the project grant ``The New Milky Way'' from the Knut and Alice Wallenberg Foundation. 
The computations were performed on resources provided by the Swedish National Infrastructure for Computing (SNIC) at HPC2N under project SNIC2015-1-309 and at UPPMAX under project p2013234.
\end{acknowledgements}

\bibliography{nordlander_etal_smss0313}

\end{document}